# Fast approximate Bayesian inference of HIV indicators using PCA adaptive Gauss-Hermite quadrature

Adam Howes[a,*], Alex Stringer[b], Seth R. Flaxman[c], Jeffrey W. Imai-Eaton[d]

[a]*Department of Mathematics, Imperial College London*
[b]*Department of Statistics and Actuarial Science, University of Waterloo*
[c]*Department of Computer Science, University of Oxford*
[d]*Harvard T.H. Chan School of Public Health, Harvard University*

**Abstract**

Naomi is a spatial evidence synthesis model used to produce district-level HIV epidemic indicators in sub-Saharan Africa. Multiple outcomes of policy interest, including HIV prevalence, HIV incidence, and antiretroviral therapy treatment coverage are jointly modelled using both household survey data and routinely reported health system data. The model is provided as a tool for countries to input their data to and generate estimates with during a yearly process supported by UNAIDS. Previously, inference has been conducted using empirical Bayes and a Gaussian approximation, implemented via the `TMB` R package. We propose a new inference method based on an extension of adaptive Gauss-Hermite quadrature to deal with more than 20 hyperparameters. Using data from Malawi, our method improves the accuracy of inferences for model parameters, while being substantially faster to run than Hamiltonian Monte Carlo with the No-U-Turn sampler. Our implementation leverages the existing `TMB` `C++` template for the model's log-posterior, and is compatible with any model with such a template.

## 1. Introduction

Accurate estimates of HIV indicators are crucial for mounting an effective public health response to the epidemic. Producing timely estimates at the district level, where health systems are planned and delivered, is challenging. Nationally-representative household surveys, while providing the most statistically reliable data, are costly to run and typically conducted only every five years. Furthermore, sample sizes at the district level are limited. Other data sources, such as routine health surveillance of antenatal care (ANC) clinics, are available nearer to real-time, but lack population representativeness.

The Naomi small-area estimation model (Eaton et al., 2021; Esra et al., 2024) addresses these challenges by synthesising data from multiple sources to estimate HIV indicators at a district-level, by age and sex. Modelling multiple data sources jointly mitigates the limitations of any single source and increases statistical power. Software developed for Naomi (https://naomi.unaids.org) facilitates over 35 countries inputting their data and interactively generating estimates during workshops as a part of a yearly estimates process supported by UNAIDS. Generation of estimates by country teams is an important and distinctive feature of the HIV response. Drawing on expert knowledge about the data improves the accuracy of the process and strengthens trust in the resulting estimates, creating a virtuous cycle of data quality, use, and ownership (Noor, 2022).

---

*Corresponding author
*Email addresses:* `ath19@ic.ac.uk` (Adam Howes), `alex.stringer@uwaterloo.ca` (Alex Stringer), `seth.flaxman@cs.ox.ac.uk` (Seth R. Flaxman), `jeaton@hsph.harvard.edu` (Jeffrey W. Imai-Eaton)

Naomi comprises multiple linked generalized linear mixed models (GLMMs) and presents a challenging Bayesian inference problem. The model contains hundreds of fixed and random effect parameters and over 20 hyperparameters. This is substantially more than the small number that can typically be handled by approaches like integrated nested Laplace approximations [INLA; Rue et al. (2009)]. Additionally, observations depend on multiple structured additive predictors, such that Naomi falls into the class of extended latent Gaussian models [ELGMs; Stringer et al. (2023)]. Inference must be fast and have low memory usage, to allow interactive review and iteration of model results by workshop participants. The scale of the model and features of its posterior geometry (Neal, 2003) make Markov chain Monte Carlo (MCMC) approaches are prohibitively slow. Inference should also be reliable, automatic across various country settings, and require minimal manual monitoring (e.g. of MCMC convergence).

Inference for Naomi is currently conducted using an empirical Bayes (EB) approach. A Gaussian approximation to the latent field is used via the Template Model Builder (`TMB`) (Kristensen et al., 2016) R package. `TMB` is gaining popularity due to its speed and flexibility, especially in spatial statistics (Osgood-Zimmerman and Wakefield, 2023) and via the user-friendly `glmmTMB` R package (Brooks et al., 2017). Inference in `TMB` occurs via optimisation of a `C++` template function, with the option available to use a Laplace approximation to integrate out any subset of the parameters. `TMB` uses automatic differentiation (Fournier et al., 2012; Baydin et al., 2017) to calculate the derivatives required for gradient-based numerical optimisation routines and the Laplace approximation.

Although the EB approach is fast, it does not take into account hyperparameter uncertainty in the latent field posterior, potentially resulting in underestimated posterior variances. This concern could have important practical implications for use of the estimates from the Naomi model, and motivated us to look for an approach closer to full Bayesian inference. We developed a method based on adaptive Gauss-Hermite quadrature (AGHQ) extended to handle integration over many hyperparameters. AGHQ is a quadrature method based on the theory of polynomial interpolation, and is well suited to statistical estimation problems in which the integrand is well approximated by a Gaussian multiplied by a polynomial. For example, Bilodeau et al. (2024) prove stochastic convergence rates for Bayesian posterior quantities when the normalising constant is estimated using AGHQ. However, it is not computationally feasible to use AGHQ in greater than 20 dimensions directly, as exponentially many nodes are required. Instead, we used principal components analysis (PCA) of the inverse curvature at the mode to find a subspace which explained most of the hyperparameter variance. In an application to Malawi, this resulted in a grid which was tractable as it had millions of times fewer nodes than the corresponding dense grid. Our implementation of the method makes use of the existing Naomi `TMB` template, and is immediately compatible with any model with such a template.

The remainder of this paper is organised as follows. In Section 2 we give background on latent Gaussian and extended latent Gaussian models. Section 3 outlines the version of the Naomi model that we consider in this paper. In Section 4 we review the deterministic inference method for ELGMs used by Stringer et al. (2023) based on nested application of AGHQ and the Laplace approximation, before introducing the PCA-based modification we use to enable application to Naomi. In Section 5 we evaluate the accuracy of a PCA-AGHQ grid for the simplified Naomi model fit to data from Malawi, as compared with EB and gold-standard MCMC. Finally, we discuss our conclusions, and directions for future research in Section 6.

**2. Background**

*2.1. Latent Gaussian model*

Latent Gaussian models [LGMs; Rue et al. (2009)] are three-stage hierarchical models with

$$y_i \sim p(y_i \mid \eta_i, \boldsymbol{\theta}_1), \quad i \in [n] \tag{1}$$

$$\mu_i = \mathbb{E}(y_i \mid \eta_i) = g(\eta_i), \tag{2}$$

$$\eta_i = \beta_0 + \sum_{l=1}^{p} \beta_j z_{ji} + \sum_{k=1}^{r} f_k(u_{ki}), \tag{3}$$

where $[n] = \{1, \ldots, n\}$. The response variable is $\mathbf{y} = (y)_{i \in [n]}$ with likelihood $p(\mathbf{y} \mid \boldsymbol{\eta}, \boldsymbol{\theta}_1) = \prod_{i=1}^{n} p(y_i \mid \eta_i, \boldsymbol{\theta}_1)$, where $\boldsymbol{\eta} = (\eta)_{i \in [n]}$. Each response has conditional mean $\mu_i$ with inverse link function $g : \mathbb{R} \to \mathbb{R}$ such that $\mu_i = g(\eta_i)$. The vector $\boldsymbol{\theta}_1 \in \mathbb{R}^{s_1}$, with $s_1$ assumed small, are additional parameters of the likelihood. The structured additive predictor $\eta_i$ may include an intercept $\beta_0$, linear effects $\beta_j$ of the covariates $z_{ji}$, and unknown functions $f_k(\cdot)$ of the covariates $u_{ki}$. The parameters $\beta_0$, $\{\beta_j\}$, $\{f_k(\cdot)\}$ are each assigned Gaussian priors. It is convenient to collect these parameters into a vector $\mathbf{x} \in \mathbb{R}^N$ called the latent field such that $\mathbf{x} \sim \mathcal{N}(0, \boldsymbol{Q}(\boldsymbol{\theta}_2)^-)$ where $\boldsymbol{\theta}_2 \in \mathbb{R}^{s_2}$ are further parameters, again with $s_2$ assumed small. Let $\boldsymbol{\theta} = (\boldsymbol{\theta}_1, \boldsymbol{\theta}_2) \in \mathbb{R}^s$ with $m = s_1 + s_2$ be all hyperparameters, with prior $p(\boldsymbol{\theta})$.

## 2.2. Extended latent Gaussian model

Extended latent Gaussian models [ELGMs; Stringer et al. (2023)] relax the restriction that there is a one-to-one mapping between the mean response $\boldsymbol{\mu}$ and structured additive predictor $\boldsymbol{\eta}$. Instead, the structured additive predictor is redefined as $\boldsymbol{\eta} = (\eta)_{i \in [N_n]}$, where $N_n \in \mathbb{N}$ is a function of $n$, and it is possible that $N_n \neq n$. Each mean response $\mu_i$ now depends on some subset $\mathcal{J}_i \subseteq [N_n]$ of indices of $\boldsymbol{\eta}$, with $\cup_{i=1}^{n} \mathcal{J}_i = [N_n]$ and $1 \leq |\mathcal{J}_i| \leq N_n$. The inverse link function $g$ is redefined for each observation to be a possibly many-to-one mapping $g_i : \mathbb{R}^{|\mathcal{J}_i|} \to \mathbb{R}$, such that $\mu_i = g_i(\boldsymbol{\eta}_{\mathcal{J}_i})$. Importantly, this mapping allows for the presence of non-linearity in the model. Put together, ELGMs are then of the form

$$y_i \sim p(y_i \mid \boldsymbol{\eta}_{\mathcal{J}_i}, \boldsymbol{\theta}_1), \quad i \in [n] \tag{4}$$

$$\mu_i = \mathbb{E}(y_i \mid \boldsymbol{\eta}_{\mathcal{J}_i}) = g_i(\boldsymbol{\eta}_{\mathcal{J}_i}), \tag{5}$$

$$\eta_j = \beta_0 + \sum_{l=1}^{p} \beta_j z_{ji} + \sum_{k=1}^{r} f_k(u_{ki}), \quad j \in [N_n], \tag{6}$$

with latent field and hyperparameter priors as in the LGM case.

Section 3 (e.g. Equations (8) to (11)) shows that the Naomi model involves many-to-one mappings of latent field components to the structured additive predictor for each response. Because of this, the Naomi model cannot be written as an LGM of the form described in Section 2.1, but can be written as an ELGM.

## 2.3. Template Model Builder

Template Model Builder [`TMB`; Kristensen et al. (2016)] is an R package which implements the Laplace approximation. In `TMB`, derivatives are obtained using automatic differentiation [AD; Baydin et al. (2017)]. The approach of AD is to decompose any function into a sequence of elementary operations with known derivatives. The known derivatives of the elementary operations may then be composed by repeat use of the chain rule to obtain the function's derivative. A review of AD and how it can be efficiently implemented is provided by Margossian (2019). `TMB` uses the C++ package `CppAD` (Bell, 2023) for AD [Section 3; Kristensen et al. (2016)]. The development of `TMB` was strongly inspired by the Automatic Differentiation Model Builder [ADMB; Fournier et al. (2012); Bolker et al. (2013)] project. An algorithm is used in `TMB` to automatically determine matrix sparsity structure [Section 4.2; Kristensen et al. (2016)]. The R package `Matrix` and C++ package `Eigen` are then used for sparse and dense matrix calculations. Kristensen et al. (2016) highlight the modular design philosophy of `TMB`. A review of the use of `TMB` for spatial modelling, including comparison to `R-INLA`, is provided by Osgood-Zimmerman and Wakefield (2023).

Models are specified in `TMB` using a C++ template file which evaluates $\log p(\mathbf{y}, \mathbf{x}, \boldsymbol{\theta})$. Other software packages have been developed which also use `TMB` C++ templates. The `tmbstan` R package (Monnahan and Kristensen, 2018) allows running the Hamiltonian Monte Carlo (HMC) algorithm via Stan (Stan Development Team). The `aghq` R package (Stringer, 2021) allows use of AGHQ, and AGHQ over the marginal Laplace approximation, via the `mvQuad` R package (Weiser, 2016). The `glmmTMB` R package (Brooks et al., 2017) allows specification of common GLMM models via a formula interface. It is also possible to extract the `TMB` objective function used by `glmmTMB`, which may then be passed into `aghq` or `tmbstan`. This discussion highlights the utility of `TMB` templates, emphasising that our approach, built on `TMB`, is both general and broadly applicable.

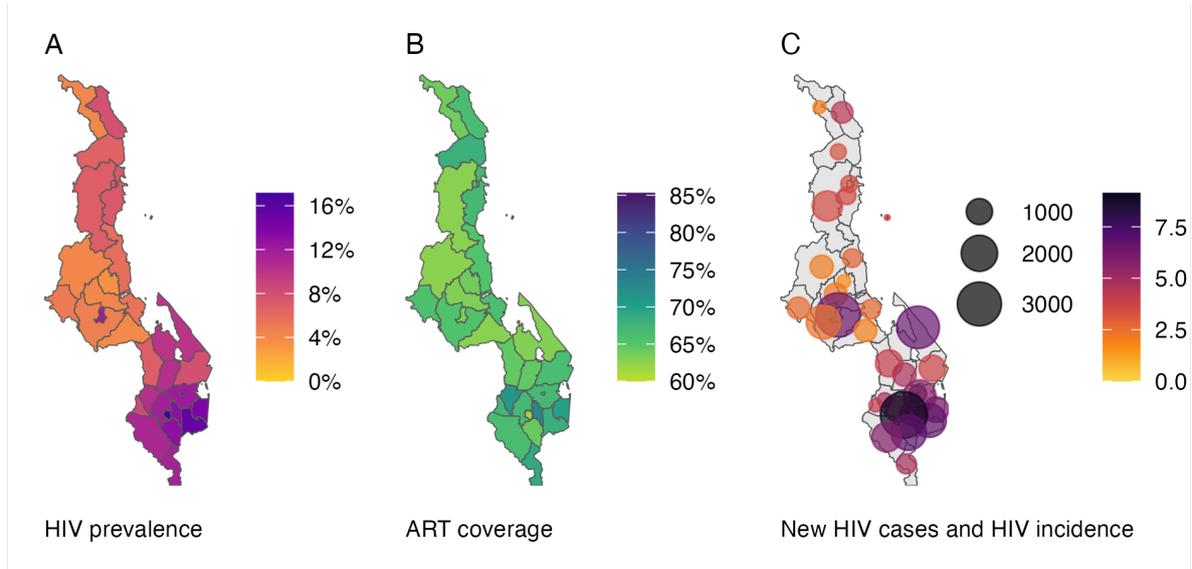

Figure 1: District-level HIV prevalence (A), ART coverage (B), and new HIV cases and HIV incidence (C) for adults 15-49 in Malawi. Inference here was conducted using and Gaussian approximation and EB via `TMB`.

## 3. Simplified Naomi model

Eaton et al. (2021) introduce a joint ELGM linking three small-area estimation models. We consider a simplified version defined only at the time of the most recent household survey with HIV testing. While this version omits nowcasting and temporal projection, as these time points involve limited inference we expect conclusions for the simplified model to be applicable to the complete model. An overview of the simplified model is given below, and a full specification is provided in Appendix 1.

### 3.1. Notation and overview

Consider a country in sub-Saharan Africa where a household survey with complex survey design has taken place. Let $x \in \mathcal{X}$ index district, $a \in \mathcal{A}$ index five-year age group, and $s \in \mathcal{S}$ index sex. For ease of notation, let $i$ index the finest district-age-sex division included in the model. Let $I \subseteq \mathcal{X} \times \mathcal{A} \times \mathcal{S}$ be a set of indices $i$ for which an aggregate observation is reported, and $\mathcal{I}$ be the set of all $I$ such that $I \in \mathcal{I}$.

Let $N_i \in \mathbb{N}$ be the known, fixed population size, $\rho_i \in [0, 1]$ be the HIV prevalence, $\alpha_i \in [0, 1]$ be the antiretroviral therapy (ART) coverage, $\kappa_i \in [0, 1]$ be the proportion who are recently infected among HIV positive persons, and $\lambda_i > 0$ be the annual HIV incidence rate.

Some observations are made at an aggregate level over a collection of strata $i$ rather than for a single $i$. Let $I \subseteq \mathcal{X} \times \mathcal{A} \times \mathcal{S}$ be a set of indices $i$ for which an aggregate observation is reported. The set of all $I$ is denoted $\mathcal{I}$ such that $I \in \mathcal{I}$.

Naomi is a joint model on the observations $\mathbf{y} = (y_I^\theta)$ for $\theta \in \{\rho, \alpha, \kappa, \rho^{\text{ANC}}, \alpha^{\text{ANC}}, N^{\text{ART}}\}$ and $I \in \mathcal{I}$. A superscript is used here to refer to components of the model (and does not refer to a power, unless made clear). The structured additive predictors contain intercept effects, age random effects, and spatial random effects which we collectively describe as the latent field $\mathbf{x}$. The latent field is controlled by hyperparamters $\boldsymbol{\theta}$ which include standard deviations, first-order autoregressive model correlation parameters, and reparameterised Besag-York-Mollie model [BYM2; Simpson et al. (2017)] proportion parameters. The number of hyperparameters is a key feature making inference challenging.

### 3.2. Household survey component

Independent logistic regression models are specified for HIV prevalence and ART coverage in the general population such that $\text{logit}(\rho_i) = \eta_i^\rho$ and $\text{logit}(\alpha_i) = \eta_i^\alpha$. HIV incidence rate is modelled by $\log(\lambda_i) = \eta_i^\lambda$ and depends on adult HIV prevalence and adult ART coverage. The structured additive predictors $\eta_i^\theta$ for $\theta \in \{\rho, \alpha, \lambda\}$ are given in Appendix 1. The proportion recently infected $\kappa_i$ is linked to HIV incidence via a reformulation of the biomarker-based incidence estimator of Kassanjee et al. (2012)

$$\kappa_i = 1 - \exp\left(-\lambda_i \cdot \frac{1-\rho_i}{\rho_i} \cdot (\Omega_T - \beta_T) - \beta_T\right), \tag{7}$$

where the mean duration of recent infection $\Omega_T$ and the proportion of long-term HIV infections misclassified as recent $\beta_T$ are strongly informed by priors for the particular survey.

These processes are each informed by household survey data. Weighted aggregate survey observations are calculated based on individual responses $\theta_j \in \{0, 1\}$ as

$$\hat{\theta}_I = \frac{\sum_j w_j \cdot \theta_j}{\sum_j w_j}. \tag{8}$$

Design weights $w_j$ for each of $\theta \in \{\rho, \alpha, \kappa\}$ are supplied by the survey provider and aim to reduce bias by decreasing possible correlation between response and recording mechanism (Meng, 2018). The index $j$ runs across all individuals in strata $i \in I$ within the relevant denominator i.e. for ART coverage, only those individuals who are HIV positive. The weighted observed number of outcomes is taken to be $y_I^\theta = m_I^\theta \cdot \hat{\theta}_I$ where

$$m_I^\theta = \frac{\left(\sum_j w_j\right)^2}{\sum_j w_j^2}, \tag{9}$$

is the Kish effective sample size [ESS; Kish (1965)]. The weighted aggregated number of outcomes are modelled using a binomial working likelihood (Chen et al., 2014) defined to operate on the reals

$$y_I^\theta \sim \text{xBin}(m_I^\theta, \theta_I) \tag{10}$$

The terms $\theta_I$ are the following weighted aggregates

$$\rho_I = \frac{\sum_{i \in I} N_i \rho_i}{\sum_{i \in I} N_i}, \quad \alpha_I = \frac{\sum_{i \in I} N_i \rho_i \alpha_i}{\sum_{i \in I} N_i \rho_i}, \quad \kappa_I = \frac{\sum_{i \in I} N_i \rho_i \kappa_i}{\sum_{i \in I} N_i \rho_i}. \tag{11}$$

*3.3. ANC testing component*

Women attending ANC clinics are routinely tested for HIV, to help prevent mother-to-child transmission. HIV prevalence $\rho_i^{\text{ANC}}$ and ART coverage $\alpha_i^{\text{ANC}}$ among pregnant women are modelled as offset from the general population indicators as follows

$$\text{logit}(\rho_i^{\text{ANC}}) = \text{logit}(\rho_i) + \eta_i^{\rho^{\text{ANC}}}, \tag{12}$$
$$\text{logit}(\alpha_i^{\text{ANC}}) = \text{logit}(\alpha_i) + \eta_i^{\alpha^{\text{ANC}}}. \tag{13}$$

These processes are informed by likelihoods specified for aggregate ANC data from the year of the most recent survey. The number of ANC clients with ascertained status to be fixed as $m_I^{\rho^{\text{ANC}}}$. The number of those with positive status $y_I^{\rho^{\text{ANC}}}$, and the number of those already on ART prior to their first ANC visit $y_I^{\alpha^{\text{ANC}}}$ are modelled using nested binomial likelihoods

$$y_I^{\rho^{\text{ANC}}} \sim \text{Bin}(m_I^{\rho^{\text{ANC}}}, \rho_I^{\text{ANC}}), \tag{14}$$
$$y_I^{\alpha^{\text{ANC}}} \sim \text{Bin}(y_I^{\rho^{\text{ANC}}}, \alpha_I^{\text{ANC}}). \tag{15}$$

As in the household survey component, weighted aggregates

$$\rho_I^{\text{ANC}} = \frac{\sum_{i \in I} \Psi_i \rho_i^{\text{ANC}}}{\sum_{i \in I} \Psi_i}, \quad \alpha_I^{\text{ANC}} = \frac{\sum_{i \in I} \Psi_i \rho_i^{\text{ANC}} \alpha_i^{\text{ANC}}}{\sum_{i \in I} \Psi_i \rho_i^{\text{ANC}}}, \tag{16}$$

are used, with $\Psi_i$ the number of pregnant women assumed here to be fixed.

*3.4. ART attendance component*

Data on attendance of ART clinics are routinely collected. These data provide helpful information about HIV prevalence and coverage of ART, but are challenging to use because people living with HIV sometimes choose to access ART services outside of the district that they reside in. These probabilities of accessing services outside the home district are modelled using multinomial logistic regressions.

Briefly, let $\gamma_{x,x'}$ be the probability that a person on ART residing in district $x$ receives ART in district $x'$, and assume $\gamma_{x,x'} = 0$ unless $x = x'$ or the two districts are neighbouring such that $x \sim x'$. The log-odds $\tilde{\gamma}_{x,x'} = \text{logit}(\gamma_{x,x'})$ are modelled using a structured additive predictor $\eta_x^{\tilde{\gamma}}$ which only depends on the home district $x$. As such, travel to each neighbouring district, for all age-sex strata, is assumed to be equally likely. The aggregate ART attendance data $y_I^{N^{\text{ART}}}$ are modelled using a Gaussian approximation to a sum of binomials. This sum is over both strata $i \in I$ and the number of ART clients travelling from district $x'$ to $x$.

## 4. Inference methods for Naomi

Section 4.1 gives the inference method for ELGMs of Stringer et al. (2023) based on nested applications of the Laplace approximation and AGHQ. In Section 4.2 we propose an extension of the method which uses PCA to facilitate inference for Naomi, which otherwise would be intractable.

*4.1. Inference for ELGMs*

The joint posterior of the parameters $(\mathbf{x}, \boldsymbol{\theta})$ given data $\mathbf{y}$ and hyperparameter prior distribution $p(\boldsymbol{\theta})$ in an ELGM is given by

$$p(\mathbf{x}, \boldsymbol{\theta} \,|\, \mathbf{y}) \propto p(\boldsymbol{\theta})|\mathbf{Q}(\boldsymbol{\theta})|^{N/2} \exp\left(-\frac{1}{2}\mathbf{x}^\top \mathbf{Q}(\boldsymbol{\theta})\mathbf{x} + \sum_{i=1}^n \log p(y_i \,|\, \mathbf{x}_{\mathcal{J}_i}, \boldsymbol{\theta})\right). \tag{17}$$

We consider approximations to the posterior marginals of each latent random variable $x_i$ and hyperparameter $\theta_j$ given by

$$p(x_i \,|\, \mathbf{y}) \approx \tilde{p}(x_i \,|\, \mathbf{y}) = \int \tilde{p}(x_i \,|\, \boldsymbol{\theta}, \mathbf{y})\tilde{p}(\boldsymbol{\theta} \,|\, \mathbf{y})\mathrm{d}\boldsymbol{\theta}, \quad i \in [N], \tag{18}$$

$$p(\theta_j \,|\, \mathbf{y}) \approx \tilde{p}(\theta_j \,|\, \mathbf{y}) = \int \tilde{p}(\boldsymbol{\theta} \,|\, \mathbf{y})\mathrm{d}\boldsymbol{\theta}_{-j} \quad j \in [m], \tag{19}$$

where the approximations $\tilde{p}(x_i \,|\, \boldsymbol{\theta}, \mathbf{y})$ and $\tilde{p}(\boldsymbol{\theta} \,|\, \mathbf{y})$ remain to be defined.

*4.1.1. Laplace approximation*

Let $\tilde{p}_{\texttt{G}}(\mathbf{x} \,|\, \boldsymbol{\theta}, \mathbf{y}) = \mathcal{N}(\mathbf{x} \,|\, \hat{\mathbf{x}}(\boldsymbol{\theta}), \hat{\mathbf{H}}(\boldsymbol{\theta})^{-1})$ be a Gaussian approximation to $p(\mathbf{x} \,|\, \boldsymbol{\theta}, \mathbf{y})$ defined by mode and precision matrix

$$\hat{\mathbf{x}}(\boldsymbol{\theta}) = \arg\max_{\mathbf{x}} \log p(\mathbf{y}, \mathbf{x}, \boldsymbol{\theta}), \tag{20}$$

$$\hat{\mathbf{H}}(\boldsymbol{\theta}) = -\frac{\partial^2}{\partial \mathbf{x} \partial \mathbf{x}^\top} \log p(\mathbf{y}, \mathbf{x}, \boldsymbol{\theta})|_{\mathbf{x}=\hat{\mathbf{x}}(\boldsymbol{\theta})}. \tag{21}$$

Then the Laplace approximation (Naylor and Smith, 1982) to $p(\boldsymbol{\theta}, \mathbf{y})$ is given by

$$\tilde{p}_{\texttt{LA}}(\boldsymbol{\theta}, \mathbf{y}) = \frac{p(\mathbf{y}, \mathbf{x}, \boldsymbol{\theta})}{\tilde{p}_{\texttt{G}}(\mathbf{x} \,|\, \boldsymbol{\theta}, \mathbf{y})}\Big|_{\mathbf{x}=\hat{\mathbf{x}}(\boldsymbol{\theta})} = \sqrt{\frac{|\hat{\mathbf{H}}(\boldsymbol{\theta})|}{(2\pi)^N}} p(\mathbf{y}, \hat{\mathbf{x}}(\boldsymbol{\theta}), \boldsymbol{\theta}). \tag{22}$$

Inference proceeds by optimising Equation (22) using a gradient-based routine to obtain $\hat{\boldsymbol{\theta}}_{\texttt{LA}} = \arg\max_{\boldsymbol{\theta}} \tilde{p}_{\texttt{LA}}(\boldsymbol{\theta}, \mathbf{y})$. Each evaluation in the optimisation requires an inner optimisation to obtain $\hat{\mathbf{x}}(\boldsymbol{\theta})$ via Equation (20). Supposing the hyperparameters are to be considered fixed, as with the `TMB` approach used currently for Naomi, then latent field joint and marginal inferences then follow directly from the Gaussian approximation $\tilde{p}_{\texttt{G}}(\mathbf{x} \,|\, \hat{\boldsymbol{\theta}}_{\texttt{LA}}, \mathbf{y})$.

*4.1.2. Adaptive Gauss-Hermite quadrature*

Let $\mathbf{z} \in \mathcal{Q}(m,k)$ be an $m$-dimensional Gauss-Hermite quadrature [GHQ; Davis and Rabinowitz (1975)] rule with $k$ nodes per dimension constructed using the product rule. In particular, $\mathcal{Q}(m,k) = \mathcal{Q}(1,k)^m$ where

$$\mathcal{Q}(1,k) = \{z : H_k(z) = 0\} \tag{23}$$

are the zeroes of the $k$th Hermite polynomial $H_k(z) = (-1)^k \exp(z^2/2)\frac{\mathrm{d}}{\mathrm{d}z^k}\exp(-z^2/2)$. The corresponding weighting function $\omega_k : \mathcal{Q}(m,k) \to \mathbb{R}$ is given by $\omega_k(\mathbf{z}) = \prod_{j=1}^{m} \omega_k(z_j)$ where $\omega_k(z) = k!/[H_{k+1}(z)]^2 \phi(z)$, and $\phi(\cdot)$ is a standard Gaussian density. GHQ is exact for functions which are a Gaussian density multiplied by a polynomial of total order no more than $2k-1$.

Let $\hat{\mathbf{H}}_{\mathtt{LA}}(\hat{\boldsymbol{\theta}}_{\mathtt{LA}}) = -\partial^2 \log p_{\mathtt{LA}}(\hat{\boldsymbol{\theta}}_{\mathtt{LA}}, \mathbf{y})$ be the curvature at the mode $\hat{\boldsymbol{\theta}}_{\mathtt{LA}}$ and $[\hat{\mathbf{H}}_{\mathtt{LA}}(\hat{\boldsymbol{\theta}}_{\mathtt{LA}})]^{-1} = \hat{\mathbf{P}}_{\mathtt{LA}}\hat{\mathbf{P}}_{\mathtt{LA}}^{\top}$ be a matrix decomposition of the inverse curvature. An adaptive Gauss-Hermite quadrature [AHGQ; Naylor and Smith (1982); Tierney and Kadane (1986)] estimate of the normalising constant $p(\mathbf{y})$ based on the Laplace approximation is given by

$$p(\mathbf{y}) \approx \int_{\boldsymbol{\theta}} \tilde{p}_{\mathtt{LA}}(\boldsymbol{\theta}, \mathbf{y}) \approx \tilde{p}_{\mathtt{AGHQ}}(\mathbf{y}) = |\hat{\mathbf{P}}_{\mathtt{LA}}| \sum_{\mathbf{z} \in \mathcal{Q}(m,k)} \tilde{p}_{\mathtt{LA}}(\hat{\mathbf{P}}_{\mathtt{LA}}\mathbf{z} + \hat{\boldsymbol{\theta}}_{\mathtt{LA}}, \mathbf{y})\omega_k(\mathbf{z}). \tag{24}$$

When $k=1$ Equation (24) corresponds to a Laplace approximation. The unadapted nodes are shifted by the mode and rotated by a matrix decomposition of the inverse curvature such that $\mathbf{z} \mapsto \hat{\mathbf{P}}_{\mathtt{LA}}\mathbf{z} + \hat{\boldsymbol{\theta}}_{\mathtt{LA}}$. Repositioning the nodes is crucial for statistical quadrature problems like ours, where the integral depends on data $\mathbf{y}$ and regions of high density are not known in advance. Two alternatives for the matrix decomposition (Jäckel, 2005) are the Cholesky decomposition $\hat{\mathbf{P}}_{\mathtt{LA}} = \hat{\mathbf{L}}_{\mathtt{LA}}$, where $\hat{\mathbf{L}}_{\mathtt{LA}}$ is lower triangular, and the spectral decomposition $\hat{\mathbf{P}}_{\mathtt{LA}} = \hat{\mathbf{E}}_{\mathtt{LA}}\hat{\boldsymbol{\Lambda}}_{\mathtt{LA}}^{1/2}$, where $\hat{\mathbf{E}}_{\mathtt{LA}} = (\hat{\mathbf{e}}_{\mathtt{LA},1}, \ldots \hat{\mathbf{e}}_{\mathtt{LA},m})$ contains the eigenvectors of $[\hat{\mathbf{H}}_{\mathtt{LA}}(\hat{\boldsymbol{\theta}}_{\mathtt{LA}})]^{-1}$ and $\hat{\boldsymbol{\Lambda}}_{\mathtt{LA}}$ is a diagonal matrix containing its eigenvalues $(\hat{\lambda}_{\mathtt{LA},1}, \ldots, \hat{\lambda}_{\mathtt{LA},m})$. This estimate may be used to normalise the Laplace approximation

$$\tilde{p}_{\mathtt{LA}}(\boldsymbol{\theta} \mid \mathbf{y}) = \frac{\tilde{p}_{\mathtt{LA}}(\boldsymbol{\theta}, \mathbf{y})}{\tilde{p}_{\mathtt{AGHQ}}(\mathbf{y})}. \tag{25}$$

To obtain inferences for the latent field (Equation (18)) we reuse the adapted nodes and weights (Rue et al., 2009; Stringer et al., 2023)

$$\tilde{p}(\mathbf{x} \mid \mathbf{y}) = |\hat{\mathbf{P}}_{\mathtt{LA}}| \sum_{\mathbf{z} \in \mathcal{Q}(m,k)} \tilde{p}_{\mathtt{G}}(\mathbf{x} \mid \hat{\mathbf{P}}_{\mathtt{LA}}\mathbf{z} + \hat{\boldsymbol{\theta}}_{\mathtt{LA}}, \mathbf{y})\tilde{p}_{\mathtt{LA}}(\hat{\mathbf{P}}_{\mathtt{LA}}\mathbf{z} + \hat{\boldsymbol{\theta}}_{\mathtt{LA}} \mid \mathbf{y})\omega_k(\mathbf{z}). \tag{26}$$

Samples from this mixture of Gaussians may be obtained by drawing a node $\mathbf{z}$ with multinomial probabilities $\lambda(\mathbf{z}) = |\hat{\mathbf{P}}_{\mathtt{LA}}|p_{\mathtt{LA}}(\hat{\mathbf{P}}_{\mathtt{LA}}\mathbf{z} + \hat{\boldsymbol{\theta}}_{\mathtt{LA}} \mid \mathbf{y})\omega_k(\mathbf{z})$, then drawing from the corresponding Gaussian $\tilde{p}_{\mathtt{G}}(\mathbf{x} \mid \hat{\mathbf{P}}_{\mathtt{LA}}\mathbf{z} + \hat{\boldsymbol{\theta}}_{\mathtt{LA}}, \mathbf{y})$.

*4.2. Principal components analysis*

Use of the product rule grid described above requires $|\mathcal{Q}(m,k)| = k^m$ quadrature points which quickly becomes intractable as $m$ increases for $k > 1$. We propose to let $\mathbf{k} = (k_1, \ldots, k_m)$ be a vector of levels for each dimension of $\boldsymbol{\theta}$. Then define $\mathcal{Q}(m,\mathbf{k}) = \mathcal{Q}(1,k_1) \times \cdots \times \mathcal{Q}(1,k_m)$ as a GHQ grid with possible variable levels of size $|\mathcal{Q}(m,\mathbf{k})| = \prod_{j=1}^{m} k_j$. Let $\mathcal{Q}(m,s,k)$ correspond to $\mathcal{Q}(m,\mathbf{k})$ with choice of levels $k_j = k, j \leq s$ and $k_j = 1, j > s$ for some $s \leq m$. For example, for $m = 2$ and $s = 1$ then $\mathbf{k} = (k,1)$. In combination with use of the spectral decomposition, this choice of levels is analogous to a principal components analysis (PCA) approach to AGHQ. We describe this approach as PCA-AGHQ, with corresponding estimate of the normalising constant given by

$$\tilde{p}_{\mathtt{PCA}}(\mathbf{y}) = |\hat{\mathbf{E}}_{\mathtt{LA}}\hat{\boldsymbol{\Lambda}}_{\mathtt{LA}}^{1/2}| \sum_{\mathbf{z} \in \mathcal{Q}(m,s,k)} \tilde{p}_{\mathtt{LA}}(\hat{\mathbf{E}}_{\mathtt{LA},s}\hat{\boldsymbol{\Lambda}}_{\mathtt{LA},s}^{1/2}\mathbf{z} + \hat{\boldsymbol{\theta}}_{\mathtt{LA}}, \mathbf{y})\widetilde{\omega}(\mathbf{z}), \tag{27}$$

where $\hat{\mathbf{E}}_{\mathtt{LA},s}$ is an $m \times s$ matrix containing the first $s$ eigenvectors, $\hat{\boldsymbol{\Lambda}}_{\mathtt{LA},s}$ is the $s \times s$ diagonal matrix containing the first $s$ eigenvalues, and $\widetilde{\omega}(\mathbf{z}) = \prod_{j=1}^{s} \omega_s(z_j) \times \prod_{j=s+1}^{d} \omega_1(z_j)$. Panel C of Figure 2 illustrates PCA-AGHQ for a case when $m = 2$ and $s = 1$. As AGHQ with $k = 1$ corresponds to the Laplace approximation, PCA-AGHQ can be interpreted as performing AGHQ on the first $s$ principal components of the inverse curvature, and a Laplace approximation on the remaining $m - s$ principal components. Inference for the latent field follows analogously to Equation (26).

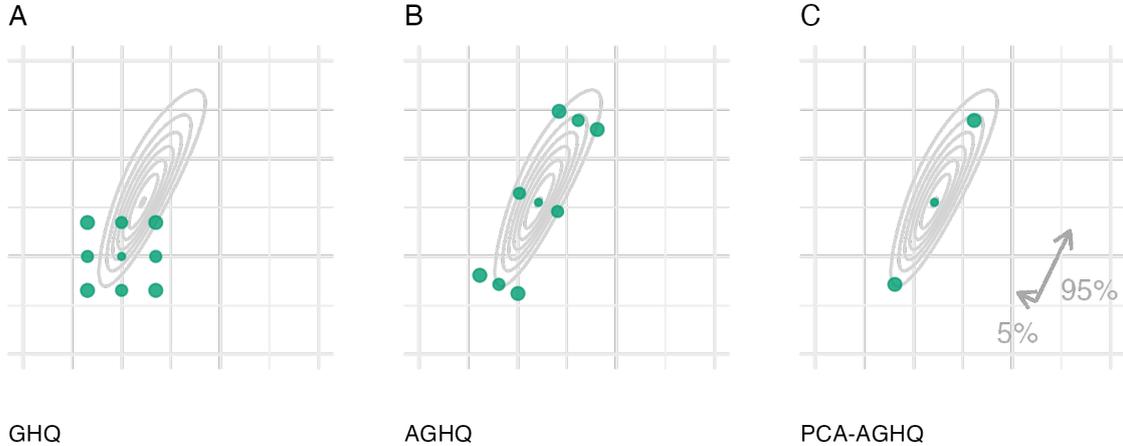

Figure 2: The Gauss-Hermite quadrature nodes $\mathbf{z} \in \mathcal{Q}(2, 3)$ for a two dimensional integral with three nodes per dimension (A). Adaption occurs based on the mode and covariance matrix of the target via the Cholesky decomposition of the inverse curvature at the mode (B). In PCA-AGHQ (C) only nodes along the first $s$ principal components are kept. Here, 95% of variation is explained by the first principal component. The integrand is $f(\boldsymbol{\theta}) = \text{sn}(0.5\theta_1, \alpha = 2) \cdot \text{sn}(0.8\theta_1 - 0.5\theta_2, \alpha = -2)$, where $\text{sn}(\cdot)$ is the standard skewnormal probability density function with shape parameter $\alpha \in \mathbb{R}$.

## 5. Application to data from Malawi

We fit the simplified Naomi model (Section 3) to data from Malawi. Malawi, which has $n = 30$ districts, has previously been used to demonstrate the Naomi model, including as a part of the `naomi` R package vignette. Three Bayesian inference approaches were considered:

1. Gaussian marginals and EB with `TMB`. This approach was previously used in production for Naomi. As short-hand, this approach is referred to as GEB.
2. Gaussian marginals and PCA-AGHQ with `TMB`. This is a novel approach. As short-hand, this approach is referred to as GPCA-AGHQ.
3. The Hamiltonian Monte Carlo algorithm No-U-Turn Sampling (NUTS) with `tmbstan` (Monnahan and Kristensen, 2018). Conditional on assessing chain convergence and suitability, inferences from NUTS represent a gold-standard.

Our goal was to determine the accuracy of the approximate methods (GEB and GPCA-AGHQ) as compared with the gold-standard (NUTS). Settings used for each inferential method are provided in Table 1, and, where relevant, discussed further below. The `TMB` C++ user-template used to specify the log-posterior was the same for each approach. The dimension of the latent field was $N = 467$ and the dimension of the hyperparameters was $m = 24$. For GEB and GPCA-AGHQ, hyperparameter and latent field samples were simulated following deterministic inference. For all methods, age-sex-district specific HIV prevalence, ART coverage and HIV incidence were simulated from the latent field and hyperparameter posteriors.

The R (R Core Team, 2021) code used to produce all results we describe below is available at `github.com/athowes/naomi-aghq`. We used `orderly` (FitzJohn et al., 2022) for reproducible research,

| Method | Software | Details |
| --- | --- | --- |
| GEB | `TMB` | 1000 samples |
| GPCA-AGHQ | `aghq` | $k = 3, s = 8$, 1000 samples |
| NUTS | `tmbstan` | 18 chains of 40000 iterations, with the first half of each chain discarded as warmup, thinned by a factor of 20, to give a total of 18000 samples. The target average acceptance probability was set to 0.95 and the maximum treedepth to 12 (Hoffman et al., 2014). |

Table 1: A summary of settings used for each inferential method.

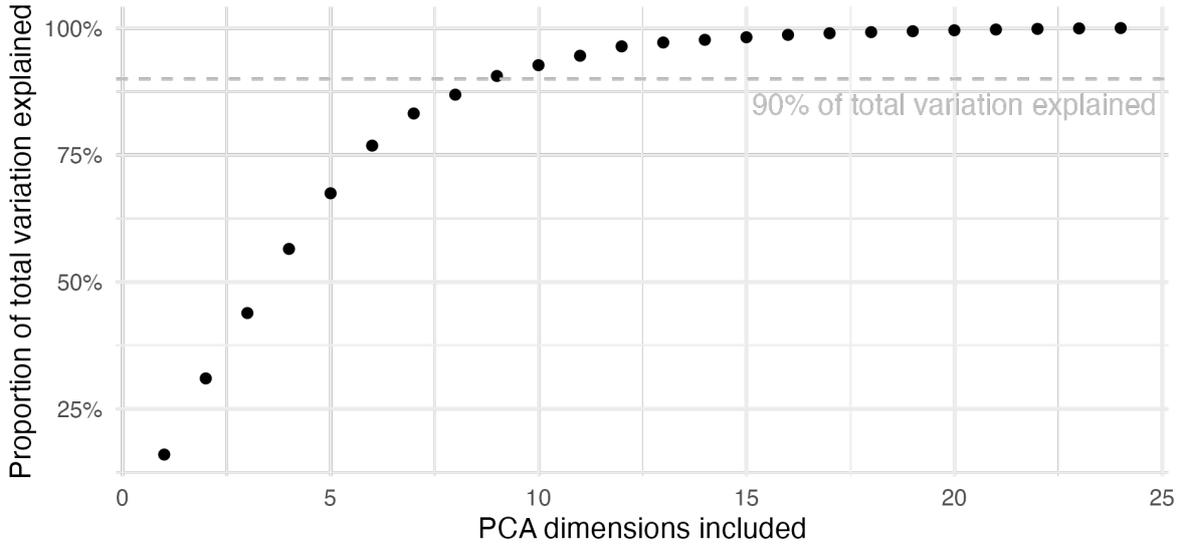

Figure 3: Under PCA, the proportion of total variation explained is given by the sum of the first $s$ eigenvalues over the sum of all eigenvalues. A typical rule-of-thumb is to include dimensions sufficient to explain 90% of total variation. In this case, for computational reasons 87% was considered sufficient.

`ggplot2` for data visualisation (Wickham, 2016) and `rticles` (Allaire et al., 2022a) to write this paper via `rmarkdown` (Allaire et al., 2022b).

*5.1. NUTS convergence and suitability*

The Naomi model was challenging to efficiently sample from using NUTS via `tmbstan`. We used 18 chains of 40000 iterations run on a high-performance computing cluster (Table 1). The lowest resulting effective sample size (ESS) was 1696 (2.5% quantile 5715, 50% quantile 14074, and 97.5% quantile 17727; Supplementary Figure 1A). The largest resulting potential scale reduction factor ($\hat{R}$) was 1.01 (2.5% quantile 1, 50% quantile 1, and 97.5% quantile 1.003; Supplementary Figure 2). Supplementary Figure 3 shows traceplots for the parameters with the lowest ESS and $\hat{R}$ respectively. Based on these diagnostics, we considered the NUTS samples to be of a high quality and suitable for use as a gold-standard.

*5.2. Use of PCA-AGHQ*

For the PCA-AGHQ quadrature grid, a Scree plot based on the spectral decomposition of $\hat{\mathbf{H}}_{\text{LA}}^{-1}$ was used to select the number of principal components to keep (Figure 3). Keeping $s = 8$ principal components was sufficient to explain 87% of total variation. The reduced rank approximation to the inverse curvature with this choice of $s$ was visually similar to the full rank matrix (Figure 4).

The principal component (PC) loadings (Supplementary Figure 8) provide interpretable information about which directions had the greatest variation. Several of the first $s = 8$ PC loadings are sums of two hyperparameters, indicating some redundancy in the hyperparameter parameterisation. This finding is supported by analysis of correlation structure in the NUTS hyperparameter posterior (Appendix 2.1).

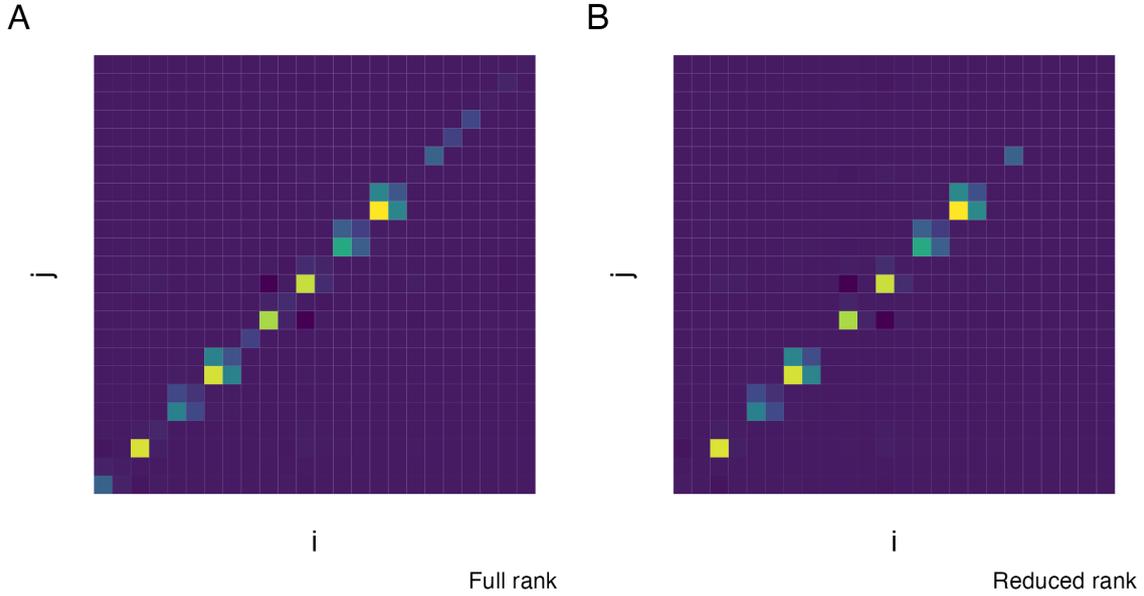

Figure 4: The full rank original covariance matrix (Panel A) was closely reproduced by its reduced rank ($s = 8$) matrix approximation (Panel B).

Projecting the $3^8 = 6561$ PCA-AGHQ quadrature nodes onto each univariate hyperparameter dimension, there was substantial variation in coverage by hyperparameter (Figure 5). Approximately 12 hyperparameters had well covered marginals: greater than the 8 naively obtained with a dense grid, but nonetheless far fewer than the full 24. Coverage was higher among hyperparameters on the logistic scale, and lower among hyperparameters on the logarithmic scale (Supplementary Figure 9). This discrepancy occurred due to logistic hyperparameters naturally having higher posterior marginal standard deviation than logarithmic hyperparameters (Supplementary Figure 10).

*5.3. Time taken comparison*

Inference with NUTS took 71 hours (across 18 cores, i.e. over 50 days of compute time) running on a high performance computing cluster. Meanwhile, inference with GPCA-AGHQ took 1.4 hours and GEB just 0.015 hours (equivalent to 0.91 minutes), both running on a laptop. The NUTS and GPCA-AGHQ algorithms can be run under a range of settings, trading off accuracy and run-time.

*5.4. Inference comparison*

Posterior inferences from GEB, GPCA-AGHQ and NUTS were compared using point estimates (Section 5.4.1), distributional quantities (Section 5.4.2), and exceedance probabilities (Section 5.4.3).

*5.4.1. Point estimates*

Latent field mean and standard deviation point estimates obtained from GPCA-AGHQ were more accurate than those from GEB (Figure 6). The root mean square error (RMSE) between posterior mean estimates from GPCA-AGHQ and NUTS (0.049) was 26% lower than that between GEB and NUTS (0.066). For the posterior standard deviation estimates, there was a substantial 70% reduction in RMSE: from 0.16 (GEB) to 0.042 (GPCA-AGHQ). Improvements in latent field estimate accuracy transferred to model outputs to a more limited and mixed extent (Supplementary Figures 12 and 13).

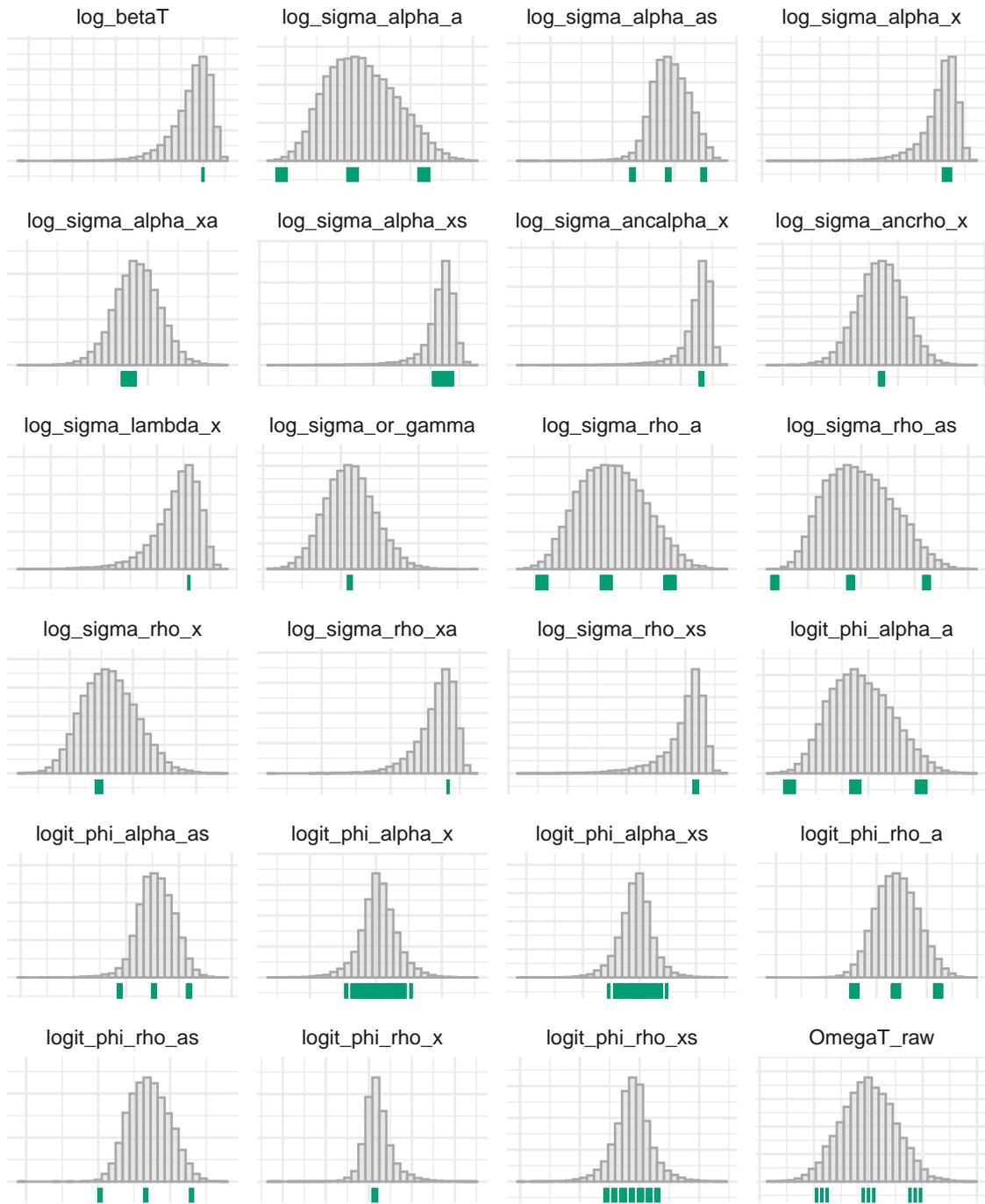

Figure 5: The grey histograms show the 24 hyperparameter marginal distributions obtained with NUTS. The green lines indicate the position of the 6561 PCA-AGHQ nodes projected onto each hyperparameter marginal. For some hyperparameters, the PCA-AGHQ nodes vary over the domain of the posterior marginal distribution, while for others they concentrate at the mode.

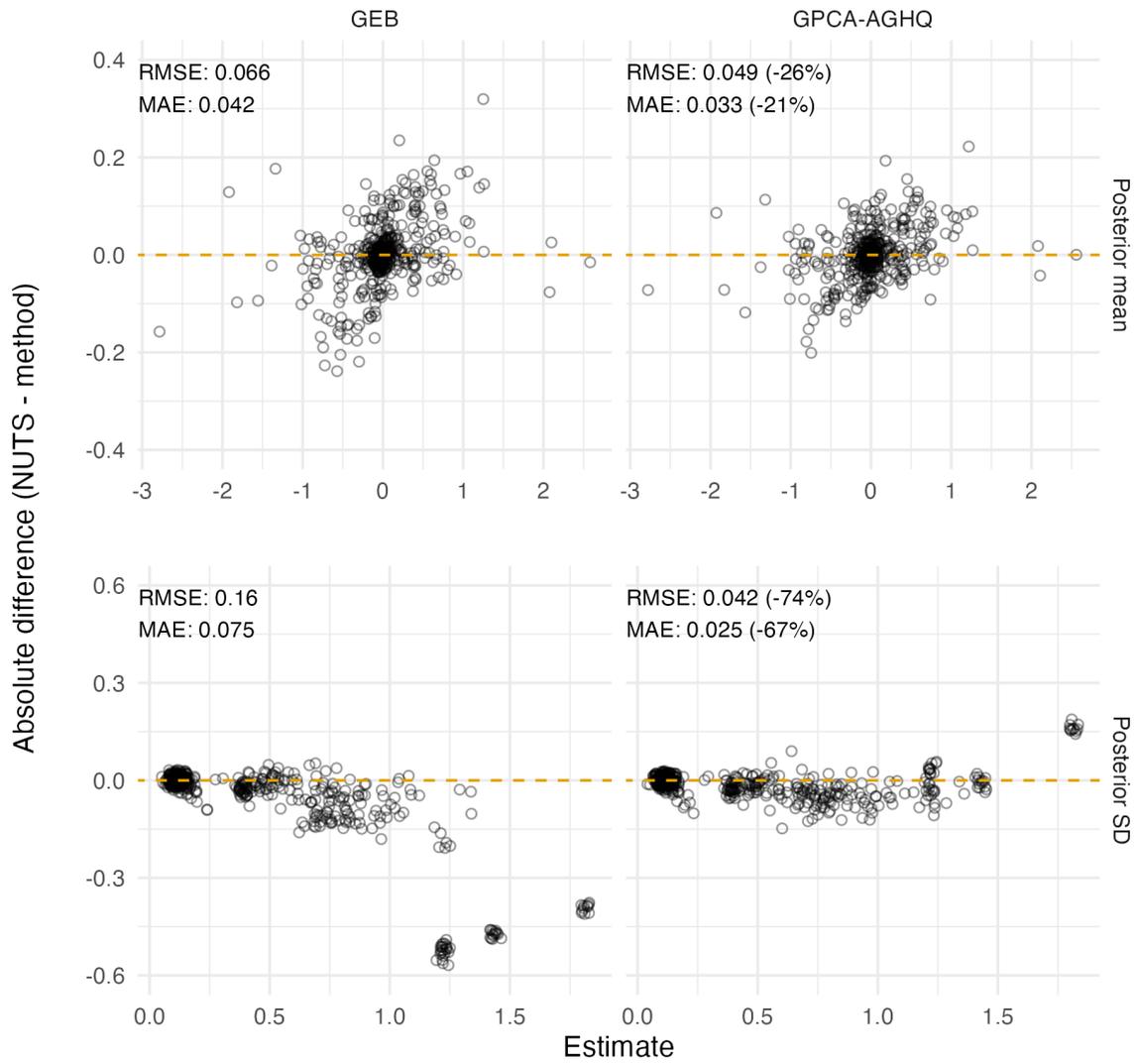

Figure 6: The latent field posterior mean and posterior standard deviation point estimates from each inference method as compared with those from NUTS. The root mean square error (RMSE) and mean absolute error (MAE) are displayed in the top left. For the posterior mean and posterior standard deviation, GPCA-AGHQ reduced RMSE and MAE as compared with GEB.

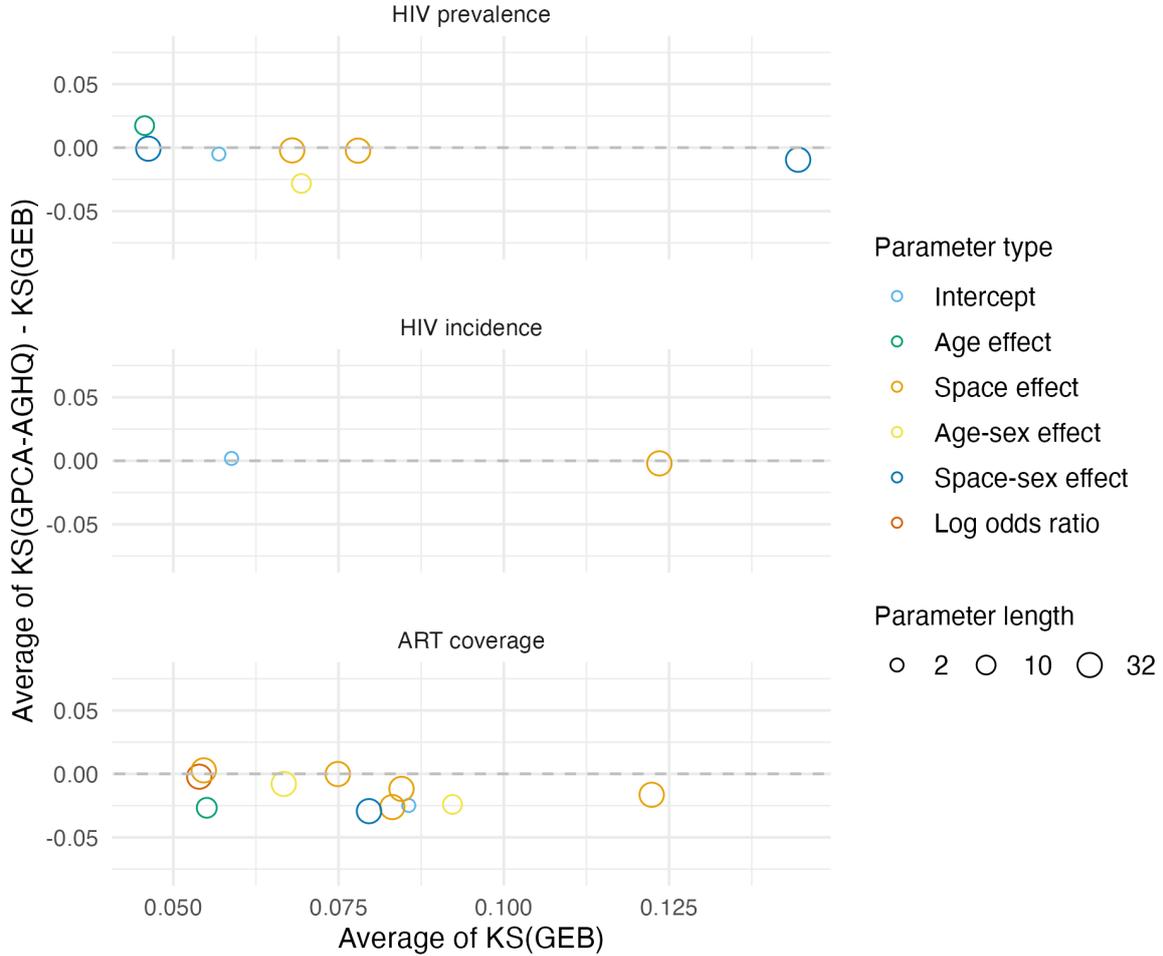

Figure 7: The average Kolmogorov-Smirnov (KS) test statistic for each latent field parameter of the Naomi model. Vectors of parameters were grouped together. For points above the dashed line at zero, performance of GEB was better. For points below the dashed line, performance of GPCA-AGHQ was better. The latent field parameters informing the ART coverage component of the model (Section 3.4) had the greatest reduction in average test statistic.

*5.4.2. Distributional quantities*

*5.4.2.1. Kolmogorov-Smirnov.* The two-sample Kolmogorov-Smirnov (KS) test statistic (Smirnov, 1948) is the maximum absolute difference between two ECDFs $F(\omega) = \frac{1}{n} \sum_{i=1}^{n} \mathbb{I}_{\phi_i \leq \omega}$. It is a relatively stringent worst-case measure of distance between empirical distributions. The average KS test statistic for GPCA-AGHQ (0.072) was 11% less than the average KS test statistic for GEB (0.081). Figure 7 shows the differences in average KS test statistic for each latent field parameter. GCPA-AGHQ showed the greatest average improvement in KS test statistic for parameters in the ART coverage component of Naomi (Figure 8). For both GEB and GPCA-AGHQ the KS test statistic for a parameter was correlated with lower NUTS ESS (Supplementary Figure 14). This correlation may be due to by difficulties estimating particular parameters for all inference methods.

*5.4.2.2. Maximum mean discrepancy.* The maximum mean discrepancy [MMD; Gretton et al. (2006)] is a measure of distance between joint distributions and can be estimated empirically using samples. Let $\Phi^1 = \{\phi_s^1\}_{s=1}^S$ and $\Phi^2 = \{\phi_s^2\}_{s=1}^S$ be two sets of joint posterior samples and $k$ be a kernel, then

$$\text{MMD}(\Phi^1, \Phi^2) = \sqrt{\frac{1}{S^2}\sum_{s,l=1}^S k(\phi_s^1, \phi_l^1) - \frac{2}{S^2}\sum_{s,l=1}^S k(\phi_s^1, \phi_l^2) + \frac{1}{S^2}\sum_{s,l=1}^S k(\phi_i^2, \phi_l^2)}. \tag{28}$$

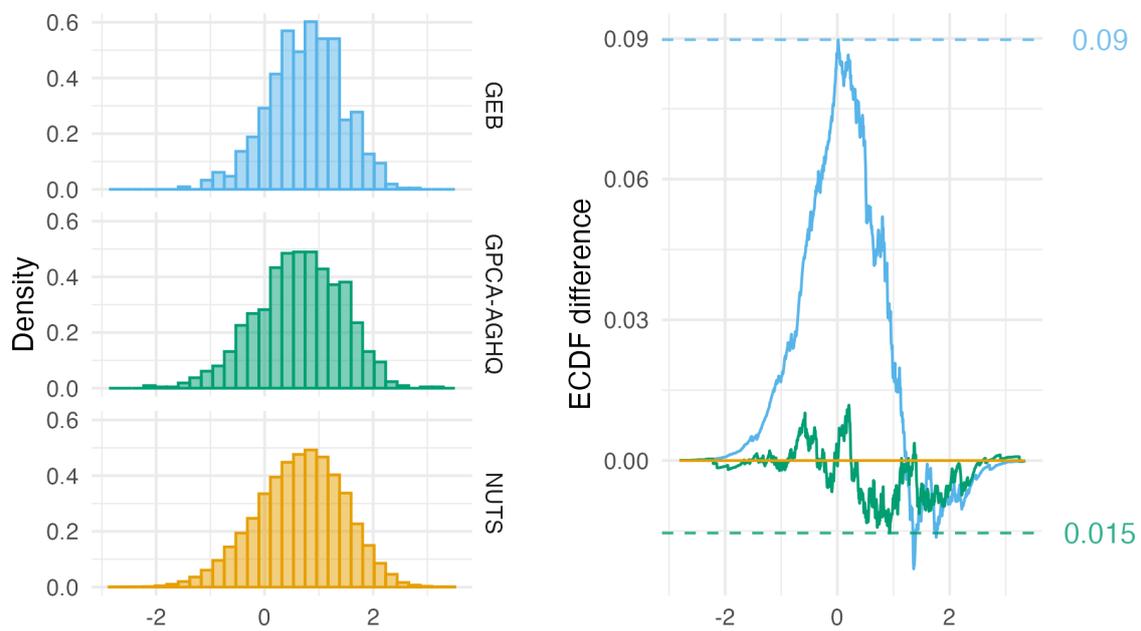

Figure 8: The parameters `us_alpha_xs` had the greatest difference in KS test statistics between GEB and GPCA-AGHQ to NUTS. This figure shows the element `us_alpha_xs[18]`. For this element, the NUTS potential scale reduction factor was 1 and effective sample size was 10700.

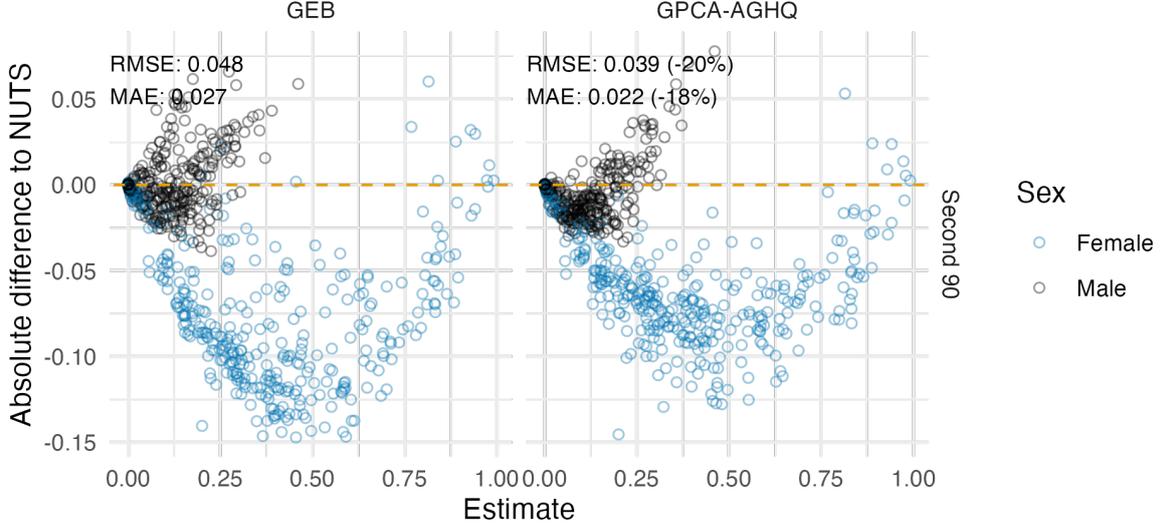

Figure 9: The probability each strata has met the second 90 (ART coverage above 81%) calculated using each inference method, as compared with NUTS. The root mean square error (RMSE) and mean absolute error (MAE) are displayed in the top left.

The kernel was set to $k(\phi^1, \phi^2) = \exp(-\sigma\|\phi^1 - \phi^2\|^2)$ with $\sigma$ estimated from data using the `kernlab` R package (Karatzoglou et al., 2019). Then, the first and third order MMD statistics for GEB were 0.078 and 0.0036 while those of GPCA-AGHQ (0.071 and 0.0024) were 9% and 30% lower.

*5.4.3. Exceedance probabilities*

As realistic use cases for Naomi model outputs, we considered the following two case-studies based on exceedance probabilities.

*5.4.3.1. Meeting the second 90.* Ambitious targets for scaling up ART treatment have been developed by UNAIDS, with the goal of ending the AIDS epidemic by 2030 (UNAIDS, 2014). Meeting the 90-90-90 fast-track target requires that 90% of people living with HIV know their status, 90% of those are on ART, and 90% of those have a suppressed viral load. Inferences from Naomi can be used to identify treatment gaps by calculating the probability that the second 90 target has been met, that is $\mathbb{P}(\alpha_i > 0.9^2 = 0.81)$ for each strata $i$. Strata probabilities of having met the second 90 target were more accurately estimated by GPCA-AGHQ than GEB (Figure 9). However, both GPCA-AGHQ and GEB had substantial error as compared to results from NUTS, particularly for women and girls. The discrepancy in accuracy by sex may be due to a more challenging posterior geometry caused by interactions between the household survey and ANC components of the model.

*5.4.3.2. Finding strata with high incidence.* Some HIV interventions are cost-effective only within high HIV incidence settings, typically defined as higher than 1% incidence per year. Inferences from Naomi can be used to calculate the probability of a strata having high incidence by evaluating $\mathbb{P}(\lambda_i > 0.01)$. GPCA-AGHQ gave more accurate estimates of the probability that a strata has high HIV incidence than GEB (Figure 10). Again, both methods had significant error as compared with NUTS.

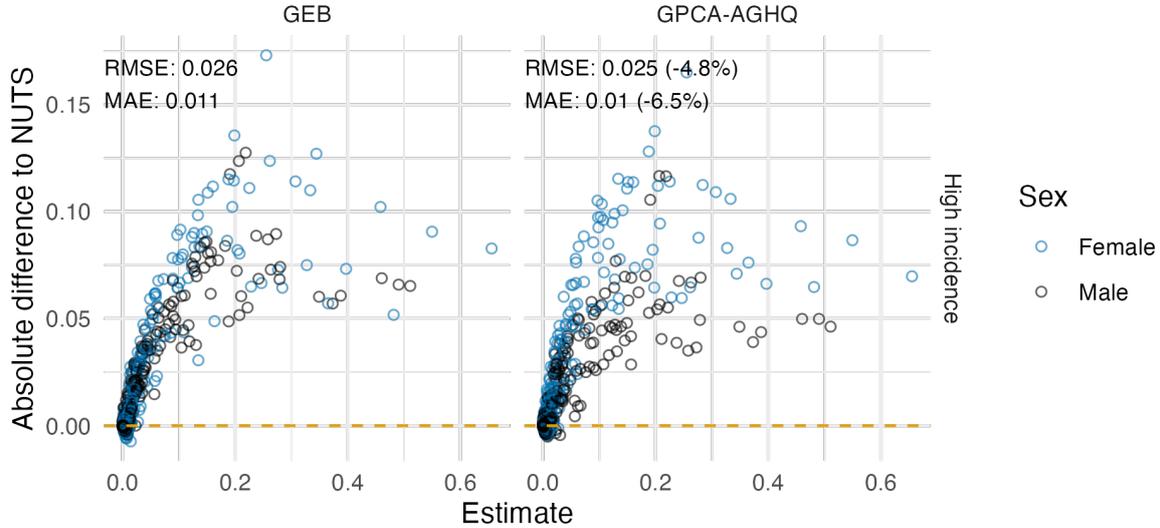

Figure 10: The probability each strata has high HIV incidence (above 1% per year) calculated using each inference method, as compared with NUTS. The root mean square error (RMSE) and mean absolute error (MAE) are displayed in the top left.

## 6. Conclusion

In the Malawi case-study, GPCA-AGHQ more accurately inferred latent field posterior marginal distributions than GEB. Benefits were particularly large for latent field posterior marginal standard deviations (Figure 6). However, the benefit of using GPCA-AGHQ for model outputs, obtained by a link function applied to a linear transformation of latent field parameters, was more limited. Although posterior exceedance probabilities based on model outputs from GPCA-AGHQ were more accurate (Figures 9 and 10), both GEB and GPCA-AGHQ showed systematic inaccuracies as compared with NUTS.

Inaccuracies in model outputs from GEB and GPCA-AGHQ do have potential to meaningfully mislead policy. As such, where it is possible to do so, gold-standard NUTS results should be computed. Although NUTS is too slow to run during a workshop, as the UNAIDS HIV estimates process occurs annually, it would be viable to compute more accurate estimates afterwards. That said, Malawi is one of the countries with the fewest number of districts. As NUTS took around three days to reach convergence in Malawi and required substantial computational resources. For larger countries, with hundreds of districts, it may be impossible to run NUTS to convergence, and faster approximate methods such as GEB and GPCA-AGHQ may be necessary.

To empower users, GPCA-AGHQ and NUTS could be added to the Naomi web interface as alternatives to GEB. Analysts would be able to quickly iterate over model options using GEB, before switching to a more accurate approach once they are happy with the results. One benefit of PCA-AGHQ is that it can be adjusted to suit the computational budget available by choice of the number of dimensions kept in the PCA $s$ and the number of points per dimension $k$. The scree plot is a well established heuristic for choosing $s$, while heuristics for choosing $k$ are less well established. Whether it is preferable for a given computational budget to increase $s$ or increase $k$ is an open question. Further strategies, such as gradually lowering $k$ over the principal components, could also be considered.

### 6.1. Suggestions for future work

#### 6.1.1. Further comparisons

Comparison to further Bayesian inference methods could prove beneficial. Four possibilities stand out as being particularly valuable. First, there exist other quadrature rules for moderate dimension, such as the Box-Wilson central composite design (Box and Wilson, 1992). It would be of interest to compare INLA with a PCA-AGHQ rule to INLA with other such quadrature rules. Second, rather than use quadrature to integrate the marginal Laplace approximation, an alternative approach is to run HMC (Monnahan and Kristensen, 2018; Margossian et al., 2020). This method can be run by using `tmbstan` and setting `laplace = TRUE`. When run to convergence, inferential error of this method would solely be due to the Laplace approximation, helping to clarify the extent to which the inferential error of INLA is attributable to the quadrature grid. Third, NUTS is not especially well suited to sampling from Gaussian latent field models like Naomi. Other MCMC algorithms, such as blocked Gibbs sampling (Geman and Geman, 1984) or slice sampling (Neal, 2003), could be considered. It may be difficult to implement such algorithms using `TMB`. Many MCMC algorithms are implemented and customisable (including, for example, the choice of block structure) within the `NIMBLE` probabilistic programming language (PPL) (de Valpine et al., 2017). Direct implementation of Naomi within the Stan PPL could alter the speed by as much as plus or minus 50% based on experiements conducted by others (Monnahan and Kristensen, 2018) and would give access to other algorithms implemented in Stan such as variational inference (Kucukelbir et al., 2015) and Pathfinder (Zhang et al., 2022). However, requiring rewriting the Naomi model log-posterior outside of `TMB` would be a substantial downside. Finally, it would be of substantial interest to implement the Naomi model using the iterative INLA method via `inlabru`. However, as `inlabru`, like `R-INLA`, is based on a formula interface, it may not be possible to do so directly.

*6.1.2. Better quadrature grids*

PCA-AGHQ is a sensible approach to allocating more computational effort to dimensions which contribute more to the integral in question. Its application to Naomi surfaced several ideas for improvement and challenges to be overcome. First, the amount of variation explained in the Hessian matrix may not be of direct interest. For the Naomi model, interest is in the effect of including each dimension on the relevant model outputs. As such, using alternative measures of importance from sensitivity analysis, such as Shapley values (Shapley et al., 1953) or Sobol indices, could be preferable. This may be achievable by applying the `ADREPORT` function to model outputs within the `TMB` template to obtain gradients. Second, use of PCA is challenging when the dimensions have different scales. For the Naomi model, logit-scale hyperparameters were systematically favoured over those on the log-scale. Third, when the quadrature rule is used within an INLA algorithm, it is more important to allocate quadrature nodes to those hyperparameter marginals which are non-Gaussian. This is because the Laplace approximation is exact when the integrand is Gaussian, so a single quadrature node is sufficient. The difficulty is knowing in advance which marginals will be non-Gaussian. This could be done if there were a cheap way to obtain posterior means, which could then be compared to posterior modes obtained using optimisation. Another approach would be to measure the fit of marginal samples from a cheap approximation, like EB. The measures of fit would have to be for marginals, ruling out approaches like PSIS (Yao et al., 2018) which operate on joint distributions. Finally, it is likely possible to achieve better performance by pruning (dropping points where the weight is below some threshold) or prerotation (before adaption, in order to e.g. achieve better coverage along the principal axis) of the quadrature grid (Jäckel, 2005).

*6.1.3. Statistical theory*

The class of functions which are integrated exactly by PCA-AGHQ remains to be shown. Theorem 1 of Stringer et al. (2023) bounds the total variation error of AGHQ, establishing convergence in probability of coverage probabilities under the approximate posterior distribution to those under the true posterior distribution. Similar theory could be established for PCA-AGHQ, or more generally AGHQ with varying levels. The challenge of connecting this theory to nested use of any quadrature rule, like GPCA-AGHQ, remains an important open question.

*6.1.4. Testing quadrature assumptions*

It may be possible to test the assumptions underlying use of AGHQ grids, enabling assessment of their suitability for specific integrals. In particular, AGHQ assumes that the integrand is well approximated by a polynomial multiplied by a Gaussian density. This assumption could be tested by fitting a model using a polynomial times Gaussian kernel, given NUTS hyperparameter samples (or even better, hyperparameter samples obtained from the Laplace NUTS approach mentioned in Section 6.1.1). This approach could be generalised to also test the suitability of PCA-AGHQ grids.

**Acknowledgements**

AH was supported by the EPSRC Centre for Doctoral Training in Modern Statistics and Statistical Machine Learning (EP/S023151/1). AH conducted part of this research while an International Visiting Graduate Student at the University of Waterloo. AH and JWE were supported by the Bill and Melinda Gates Foundation (OPP1190661, OPP1164897). AS was supported by NSERC Discovery Grant (RGPIN-2023-03331). SRF was supported by the EPSRC (EP/V002910/2). JWE was supported by UNAIDS and National Institute of Allergy and Infectious Disease of the National Institutes of Health (R01AI136664). This research was supported by the MRC Centre for Global Infectious Disease Analysis (MR/R015600/1), jointly funded by the UK Medical Research Council (MRC) and the UK Foreign, Commonwealth & Development Office (FCDO), under the MRC/FCDO Concordat program and is also part of the EDCTP2 programme supported by the European Union.

# Appendix to "Fast approximate Bayesian inference of HIV indicators using PCA adaptive Gauss-Hermite quadrature"

Adam Howes[*]   Alex Stringer[†]   Seth R. Flaxman[‡]   Jeffrey W. Imai-Eaton[§]

# Contents

---

[*]Department of Mathematics, Imperial College London
[†]Department of Statistics and Actuarial Science, University of Waterloo
[‡]Department of Computer Science, Oxford University
[§]Harvard T.H. Chan School of Public Health, Harvard University

# 1 Simplified Naomi model description

This section describes the simplified version of the Naomi model (Eaton et al. 2021) in more detail. The concise $i$ indexing is replaced by a more complete $x, s, a$ indexing. There are four sections, as follows:

1. Section 1.1 gives the process specifications, including the terms in each structured additive predictor, along with their distributions.
2. Section 1.2 gives additional details about the likelihood terms.
3. Section 1.3 gives identifiability constraints used in circumstances where incomplete data is available for the country.
4. Section 1.4 provides details of the `TMB` implementation.

## 1.1 Process specification

Table 1: The Naomi model can be conceptualised as having five processes. This table gives the number of latent field parameters and hyperparameters in each model component, where $n$ is the number of districts in the country.

|               | Model component    | Latent field | Hyperparameter |
|---------------|--------------------|--------------|----------------|
| Section 1.1.1 | HIV prevalence     | $22 + 5n$    | 9              |
| Section 1.1.2 | ART coverage       | $25 + 5n$    | 9              |
| Section 1.1.3 | HIV incidence rate | $2 + n$      | 3              |
| Section 1.1.4 | ANC testing        | $2 + 2n$     | 2              |
| Section 1.1.5 | ART attendance     | $n$          | 1              |
|               | Total              | $51 + 14n$   | 24             |

### 1.1.1 HIV prevalence

HIV prevalence $\rho_{x,s,a} \in [0,1]$ was modelled on the logit scale using the structured additive predictor

$$\text{logit}(\rho_{x,s,a}) = \beta_0^\rho + \beta_S^{\rho, s=\text{M}} + \mathbf{u}_a^\rho + \mathbf{u}_a^{\rho, s=\text{M}} + \mathbf{u}_x^\rho + \mathbf{u}_x^{\rho, s=\text{M}} + \mathbf{u}_x^{\rho, a<15} + \boldsymbol{\eta}_{R_x, s, a}^\rho. \tag{1}$$

Table 2 provides a description of the terms included in Equation (1). Independent half-normal prior distributions were chosen for the five standard deviation terms

$$\{\sigma_A^\rho, \sigma_{AS}^\rho, \sigma_X^\rho, \sigma_{XS}^\rho, \sigma_{XA}^\rho\} \sim \mathcal{N}^+(0, 2.5), \tag{2}$$

independent uniform prior distributions for the two AR1 correlation parameters

$$\{\phi_A^\rho, \phi_{AS}^\rho\} \sim \mathcal{U}(-1, 1), \tag{3}$$

and independent beta prior distributions for the two BYM2 proportion parameters

$$\{\phi_X^\rho, \phi_{XS}^\rho\} \sim \text{Beta}(0.5, 0.5). \tag{4}$$

Table 2: Each term in Equation (1) together with, where applicable, its prior distribution and a written description of its role.

| Term | Distribution | Description |
|------|--------------|-------------|
| $\beta_0^\rho$ | $\mathcal{N}(0, 5)$ | Intercept |
| $\beta_s^{\rho, s=\text{M}}$ | $\mathcal{N}(0, 5)$ | The difference in logit prevalence for men compared to women |

| Term | Distribution | Description |
|---|---|---|
| $\mathbf{u}_a^\rho$ | $\text{AR1}(\sigma_A^\rho, \phi_A^\rho)$ | Age random effects for women |
| $\mathbf{u}_a^{\rho,s=\text{M}}$ | $\text{AR1}(\sigma_{AS}^\rho, \phi_{AS}^\rho)$ | Age random effects for the difference in logit prevalence for men compared to women age $a$ |
| $\mathbf{u}_x^\rho$ | $\text{BYM2}(\sigma_X^\rho, \phi_X^\rho)$ | Spatial random effects for women |
| $\mathbf{u}_x^{\rho,s=\text{M}}$ | $\text{BYM2}(\sigma_{XS}^\rho, \phi_{XS}^\rho)$ | Spatial random effects for the difference in logit prevalence for men compared to women in district $x$ |
| $\mathbf{u}_x^{\rho,a<15}$ | $\text{ICAR}(\sigma_{XA}^\rho)$ | Spatial random effects for the difference in logit paediatric prevalence to adult women prevalence in district $x$ |
| $\boldsymbol{\eta}_{R_x,s,a}^\rho$ | – | Fixed offsets specifying assumed odds ratios for prevalence outside the age ranges for which data were available. Calculated from Spectrum model (Stover et al. 2019) outputs for region $R_x$ |

### 1.1.2 ART coverage

ART coverage $\alpha_{x,s,a} \in [0,1]$ was modelled on the logit scale using the structured additive predictor

$$\text{logit}(\alpha_{x,s,a}) = \beta_0^\alpha + \beta_S^{\alpha,s=\text{M}} + \mathbf{u}_a^\alpha + \mathbf{u}_a^{\alpha,s=\text{M}} + \mathbf{u}_x^\alpha + \mathbf{u}_x^{\alpha,s=\text{M}} + \mathbf{u}_x^{\alpha,a<15} + \boldsymbol{\eta}_{R_x,s,a}^\alpha, \qquad (5)$$

with terms and prior distributions analogous to the HIV prevalence process model in Section 1.1.1 above.

### 1.1.3 HIV incidence rate

HIV incidence rate $\lambda_{x,s,a} > 0$ was modelled on the log scale using the structured additive predictor

$$\log(\lambda_{x,s,a}) = \beta_0^\lambda + \beta_S^{\lambda,s=\text{M}} + \log(\rho_x^{15\text{-}49}) + \log(1 - \omega \cdot \alpha_x^{15\text{-}49}) + \mathbf{u}_x^\lambda + \boldsymbol{\eta}_{R_x,s,a}^\lambda. \qquad (6)$$

Table 3 provides a description of the terms included in Equation (6).

Table 3: Each term in Equation (6) together with, where applicable, its prior distribution and a written description of its role.

| Term | Distribution | Description |
|---|---|---|
| $\beta_0^\lambda$ | $\mathcal{N}(0,5)$ | Intercept term proportional to the average HIV transmission rate for untreated HIV positive adults |
| $\beta_S^{\lambda,s=\text{M}}$ | $\mathcal{N}(0,5)$ | The log incidence rate ratio for men compared to women |
| $\rho_x^{15\text{-}49}$ | – | The HIV prevalence among adults 15-49 in district $x$ calculated by aggregating age-specific HIV prevalences |
| $\alpha_x^{15\text{-}49}$ | – | The ART coverage among adults 15-49 in district $x$ calculated by aggregating age-specific ART coverages |
| $\omega = 0.7$ | – | Average reduction in HIV transmission rate per increase in population ART coverage fixed based on inputs to the Estimation and Projection Package (EPP) model |
| $\mathbf{u}_x^\lambda$ | $\mathcal{N}(0, \sigma^\lambda)$ | IID spatial random effects with $\sigma^\lambda \sim \mathcal{N}^+(0,1)$ |
| $\boldsymbol{\eta}_{R_x,s,a}^\lambda$ | – | Fixed log incidence rate ratios by sex and age group calculated from Spectrum model outputs for region $R_x$ |

The proportion who were recently infected among HIV positive persons $\kappa_{x,s,a} \in [0,1]$ was modelled as

$$\kappa_{x,s,a} = 1 - \exp\left(-\lambda_{x,s,a} \cdot \frac{1 - \rho_{x,s,a}}{\rho_{x,s,a}} \cdot (\Omega_T - \beta_T) - \beta_T\right), \qquad (7)$$

where $\Omega_T \sim \mathcal{N}(\Omega_{T_0}, \sigma^{\Omega_T})$ is the mean duration of recent infection, and $\beta_T \sim \mathcal{N}^+(\beta_{T_0}, \sigma^{\beta_T})$ is the false recent ratio. The prior distribution for $\Omega_T$ was informed by the characteristics of the recent infection testing algorithm. For PHIA surveys this was $\Omega_{T_0} = 130$ days and $\sigma^{\Omega_T} = 6.12$ days. For PHIA surveys there was assumed to be no false recency, such that $\beta_{T_0} = 0.0$, $\sigma^{\beta_T} = 0.0$, and $\beta_T = 0$.

### 1.1.4 ANC testing

HIV prevalence $\rho_{x,a}^{\text{ANC}}$ and ART coverage $\alpha_{x,a}^{\text{ANC}}$ among pregnant women were modelled as being offset on the logit scale from the corresponding district-age indicators $\rho_{x,F,a}$ and $\alpha_{x,F,a}$ according to

$$\text{logit}(\rho_{x,a}^{\text{ANC}}) = \text{logit}(\rho_{x,F,a}) + \beta^{\rho^{\text{ANC}}} + \mathbf{u}_x^{\rho^{\text{ANC}}} + \boldsymbol{\eta}_{R_x,a}^{\rho^{\text{ANC}}}, \tag{8}$$

$$\text{logit}(\alpha_{x,a}^{\text{ANC}}) = \text{logit}(\alpha_{x,F,a}) + \beta^{\alpha^{\text{ANC}}} + \mathbf{u}_x^{\alpha^{\text{ANC}}} + \boldsymbol{\eta}_{R_x,a}^{\alpha^{\text{ANC}}}. \tag{9}$$

Table 4 provides a description of the terms included in Equation (8) and Equation (9).

Table 4: Each term in Equations (8) and (9) together with (where applicable) its prior distribution and a written description of its role. The notation $\theta$ is used as stand in for $\theta \in \{\rho, \alpha\}$.

| Term | Distribution | Description |
| --- | --- | --- |
| $\beta^{\theta^{\text{ANC}}}$ | $\mathcal{N}(0, 5)$ | Intercept giving the average difference between population and ANC outcomes |
| $\mathbf{u}_x^{\theta^{\text{ANC}}}$ | $\mathcal{N}(0, \sigma_X^{\theta^{\text{ANC}}})$ | IID district random effects with $\sigma_X^{\theta^{\text{ANC}}} \sim \mathcal{N}^+(0, 1)$ |
| $\boldsymbol{\eta}_{R_x,a}^{\theta^{\text{ANC}}}$ | – | Offsets for the log fertility rate ratios for HIV positive women compared to HIV negative women and for women on ART to HIV positive women not on ART, calculated from Spectrum model outputs for region $R_x$ |

In the full Naomi model, for adult women 15-49 the number of ANC clients $\Psi_{x,a} > 0$ were modelled as

$$\log(\Psi_{x,a}) = \log(N_{x,F,a}) + \psi_{R_x,a} + \beta^\psi + \mathbf{u}_x^\psi, \tag{10}$$

where $N_{x,F,a}$ are the female population sizes, $\psi_{R_x,a}$ are fixed age-sex fertility ratios in Spectrum region $R_x$, $\beta^\psi$ are log rate ratios for the number of ANC clients relative to the predicted fertility, and $\mathbf{u}_x^\psi \sim \mathcal{N}(0, \sigma^\psi)$ are district random effects. Here these terms are fixed to $\beta^\psi = 0$ and $\mathbf{u}_x^\psi = \mathbf{0}$ such that $\Psi_{x,a}$ are simply constants.

### 1.1.5 ART attendance

Let $\gamma_{x,x'} \in [0, 1]$ be the probability that a person on ART residing in district $x$ receives ART in district $x'$. Assume that $\gamma_{x,x'} = 0$ for $x \notin \{x, \text{ne}(x)\}$ such that individuals seek treatment only in their residing district or its neighbours $\text{ne}(x) = \{x' : x' \sim x\}$, where $\sim$ is an adjacency relation, and $\sum_{x' \in \{x,\text{ne}(x)\}} \gamma_{x,x'} = 1$.

The probabilities $\gamma_{x,x'}$ for $x \sim x'$ were modelled using multinomial logistic regression model, based on the log-odds ratios

$$\tilde{\gamma}_{x,x'} = \log\left(\frac{\gamma_{x,x'}}{1 - \gamma_{x,x'}}\right) = \tilde{\gamma}_0 + \mathbf{u}_x^{\tilde{\gamma}}. \tag{11}$$

Table 5 provides a description of the terms included in Equation (11). Fixing $\tilde{\gamma}_{x,x} = 0$ then the multinomial probabilities may be recovered using the softmax

$$\gamma_{x,x'} = \frac{\exp(\tilde{\gamma}_{x,x'})}{\sum_{x^\star \in \{x,\text{ne}(x)\}} \exp(\tilde{\gamma}_{x,x^\star})}. \tag{12}$$

Table 5: Each term in Equation (11) together with, where applicable, its prior distribution and a written description of its role. As no terms include $x'$, $\gamma_{x,x'}$ is only a function of $x$.

| Term | Distribution | Description |
| --- | --- | --- |
| $\tilde{\gamma}_0$ | – | Fixed intercept $\tilde{\gamma}_0 = -4$. Implies a prior mean on $\gamma_{x,x'}$ of 1.8%, such that a-priori $(100 - 1.8 \times \text{ne}(x))\%$ of ART clients in district $x$ obtain treatment in their home district |
| $\mathbf{u}_x^{\tilde{\gamma}}$ | $\mathcal{N}(0, \sigma_X^{\tilde{\gamma}})$ | District random effects, with $\sigma_X^{\tilde{\gamma}} \sim \mathcal{N}^+(0, 2.5)$ |

## 1.2 Additional likelihood specification

Additional useful details regarding Naomi's likelihood specification are provided here.

### 1.2.1 Household survey data

The generalised binomial $y \sim \text{xBin}(m, p)$ is defined for $y, m \in \mathbb{R}^+$ with $y \leq m$ such that

$$\log p(y) = \log \Gamma(m + 1) - \log \Gamma(y + 1) \tag{13}$$
$$- \log \Gamma(m - y + 1) + y \log p + (m - y) \log(1 - p), \tag{14}$$

where the gamma function $\Gamma$ is such that $\forall n \in \mathbb{N}, \Gamma(n) = (n-1)!$.

## 1.3 Identifiability constraints

If data are missing, some parameters are fixed to default values to help with identifiability. In particular:

1. If survey data on HIV prevalence or ART coverage by age and sex are not available then $\mathbf{u}_a^\theta = 0$ and $\mathbf{u}_{a,s=\text{M}}^\theta = 0$. In this case, the average age-sex pattern from the Spectrum is used. For the Malawi case-study, HIV prevalence and ART coverage data are not available for those aged 65+. As a result, there are $|\{0\text{-}4, \ldots, 50\text{-}54\}| = 13$ age groups included for the age random effects.
2. If no ART data, either survey or ART programme, are available but data on ART coverage among ANC clients are available, the level of ART coverage is not identifiable, but spatial variation is identifiable. In this instance, overall ART coverage is determined by the Spectrum offset, and only area random effects are estimated such that
$$\text{logit}(\alpha_{x,s,a}) = \mathbf{u}_x^\alpha + \boldsymbol{\eta}_{R_x,s,a}^\alpha. \tag{15}$$
3. If survey data on recent HIV infection are not included in the model, then $\beta_0^\lambda = \beta_S^{\lambda,s=\text{M}} = 0$ and $\mathbf{u}_x^\lambda = \mathbf{0}$. The sex ratio for HIV incidence is determined by the sex incidence rate ratio from Spectrum, and the incidence rate in all districts is modelled assuming the same average HIV transmission rate for untreated adults, but varies according to district-level estimates of HIV prevalence and ART coverage.

## 1.4 Implementation

The `TMB` C++ code for the negative log-posterior of the simplified Naomi model is available from https://github.com/athowes/naomi-aghq. Table 6 maps the mathematical notation used in Section 1 to the variable names used in the `TMB` code for all hyperparameters and latent field parameters. For further reference on the `TMB` software see Kristensen (2021).

Table 6: Correspondence between the variable name used in the Naomi `TMB` template and the mathematical notation used in Appendix 1. The parameter type, either a hyperparameter or element of the latent field, is also given. All of the parameters are defined on the real-scale in some dimension. In the final three columns ($\rho$, $\alpha$, and $\lambda$) indication is given as to which component of the model the parameter is primarily used in.

| Variable name | Notation | Type | Domain | $\rho$ | $\alpha$ | $\lambda$ |
|---|---|---|---|---|---|---|
| `logit_phi_rho_x` | $\text{logit}(\phi_X^\rho)$ | Hyper | $\mathbb{R}$ | Yes | | |
| `log_sigma_rho_x` | $\log(\sigma_X^\rho)$ | Hyper | $\mathbb{R}$ | Yes | | |
| `logit_phi_rho_xs` | $\text{logit}(\phi_{XS}^\rho)$ | Hyper | $\mathbb{R}$ | Yes | | |
| `log_sigma_rho_xs` | $\log(\sigma_{XS}^\rho)$ | Hyper | $\mathbb{R}$ | Yes | | |
| `logit_phi_rho_a` | $\text{logit}(\phi_A^\rho)$ | Hyper | $\mathbb{R}$ | Yes | | |
| `log_sigma_rho_a` | $\log(\sigma_A^\rho)$ | Hyper | $\mathbb{R}$ | Yes | | |
| `logit_phi_rho_as` | $\text{logit}(\phi_{AS}^\rho)$ | Hyper | $\mathbb{R}$ | Yes | | |
| `log_sigma_rho_as` | $\log(\sigma_{AS}^\rho)$ | Hyper | $\mathbb{R}$ | Yes | | |
| `log_sigma_rho_xa` | $\log(\sigma_{XA}^\rho)$ | Hyper | $\mathbb{R}$ | Yes | | |
| `logit_phi_alpha_x` | $\text{logit}(\phi_X^\alpha)$ | Hyper | $\mathbb{R}$ | | Yes | |
| `log_sigma_alpha_x` | $\log(\sigma_X^\alpha)$ | Hyper | $\mathbb{R}$ | | Yes | |
| `logit_phi_alpha_xs` | $\text{logit}(\phi_{XS}^\alpha)$ | Hyper | $\mathbb{R}$ | | Yes | |
| `log_sigma_alpha_xs` | $\log(\sigma_{XS}^\alpha)$ | Hyper | $\mathbb{R}$ | | Yes | |
| `logit_phi_alpha_a` | $\text{logit}(\phi_A^\alpha)$ | Hyper | $\mathbb{R}$ | | Yes | |
| `log_sigma_alpha_a` | $\log(\sigma_A^\alpha)$ | Hyper | $\mathbb{R}$ | | Yes | |
| `logit_phi_alpha_as` | $\text{logit}(\phi_{AS}^\alpha)$ | Hyper | $\mathbb{R}$ | | Yes | |
| `log_sigma_alpha_as` | $\log(\sigma_{AS}^\alpha)$ | Hyper | $\mathbb{R}$ | | Yes | |
| `log_sigma_alpha_xa` | $\log(\sigma_{XA}^\alpha)$ | Hyper | $\mathbb{R}$ | | Yes | |
| `OmegaT_raw` | $\Omega_T$ | Hyper | $\mathbb{R}$ | | | Yes |
| `log_betaT` | $\log(\beta_T)$ | Hyper | $\mathbb{R}$ | | | Yes |
| `log_sigma_lambda_x` | $\log(\sigma^\lambda)$ | Hyper | $\mathbb{R}$ | | | Yes |
| `log_sigma_ancrho_x` | $\log(\sigma_X^{\rho^{\text{ANC}}})$ | Hyper | $\mathbb{R}$ | | Yes | |
| `log_sigma_ancalpha_x` | $\log(\sigma_X^{\alpha^{\text{ANC}}})$ | Hyper | $\mathbb{R}$ | | Yes | |
| `log_sigma_or_gamma` | $\log(\sigma_X^{\tilde{\gamma}})$ | Hyper | $\mathbb{R}$ | | | |
| `beta_rho` | $(\beta_0^\rho, \beta_s^{\rho, s=\text{M}})$ | Latent | $\mathbb{R}^2$ | Yes | | |
| `beta_alpha` | $(\beta_0^\alpha, \beta_S^{\alpha, s=\text{M}})$ | Latent | $\mathbb{R}^2$ | | Yes | |
| `beta_lambda` | $(\beta_0^\lambda, \beta_S^{\lambda, s=\text{M}})$ | Latent | $\mathbb{R}^2$ | | | Yes |
| `beta_anc_rho` | $\beta^{\rho^{\text{ANC}}}$ | Latent | $\mathbb{R}$ | | Yes | |
| `beta_anc_alpha` | $\beta^{\alpha^{\text{ANC}}}$ | Latent | $\mathbb{R}$ | | Yes | |
| `u_rho_x` | $\mathbf{w}_x^\rho$ | Latent | $\mathbb{R}^n$ | Yes | | |
| `us_rho_x` | $\mathbf{v}_x^\rho$ | Latent | $\mathbb{R}^n$ | Yes | | |
| `u_rho_xs` | $\mathbf{w}_x^{\rho, s=\text{M}}$ | Latent | $\mathbb{R}^n$ | Yes | | |
| `us_rho_xs` | $\mathbf{v}_x^{\rho, s=\text{M}}$ | Latent | $\mathbb{R}^n$ | Yes | | |
| `u_rho_a` | $\mathbf{u}_a^\rho$ | Latent | $\mathbb{R}^{10}$ | Yes | | |
| `u_rho_as` | $\mathbf{u}_a^{\rho, s=\text{M}}$ | Latent | $\mathbb{R}^{10}$ | Yes | | |
| `u_rho_xa` | $\mathbf{u}_x^{\rho, a<15}$ | Latent | $\mathbb{R}^n$ | Yes | | |
| `u_alpha_x` | $\mathbf{w}_x^\alpha$ | Latent | $\mathbb{R}^n$ | | Yes | |
| `us_alpha_x` | $\mathbf{v}_x^\alpha$ | Latent | $\mathbb{R}^n$ | | Yes | |
| `u_alpha_xs` | $\mathbf{w}_x^{\alpha, s=\text{M}}$ | Latent | $\mathbb{R}^n$ | | Yes | |
| `us_alpha_xs` | $\mathbf{v}_x^{\alpha, s=\text{M}}$ | Latent | $\mathbb{R}^n$ | | Yes | |
| `u_alpha_a` | $\mathbf{u}_a^\alpha$ | Latent | $\mathbb{R}^{13}$ | | Yes | |
| `u_alpha_as` | $\mathbf{u}_a^{\alpha, s=\text{M}}$ | Latent | $\mathbb{R}^{10}$ | | Yes | |
| `u_alpha_xa` | $\mathbf{u}_x^{\alpha, a<15}$ | Latent | $\mathbb{R}^n$ | | Yes | |

| Variable name | Notation | Type | Domain | $\rho$ | $\alpha$ | $\lambda$ |
|---|---|---|---|---|---|---|
| `ui_lambda_x` | $\mathbf{u}_x^{\lambda}$ | Latent | $\mathbb{R}^n$ | | | Yes |
| `ui_anc_rho_x` | $\mathbf{u}_x^{\rho^{\text{ANC}}}$ | Latent | $\mathbb{R}^n$ | Yes | | |
| `ui_anc_alpha_x` | $\mathbf{u}_x^{\alpha^{\text{ANC}}}$ | Latent | $\mathbb{R}^n$ | | Yes | |
| `log_or_gamma` | $\mathbf{u}_x^{\tilde{\gamma}}$ | Latent | $\mathbb{R}^n$ | | | |

## 2 NUTS convergence and suitability

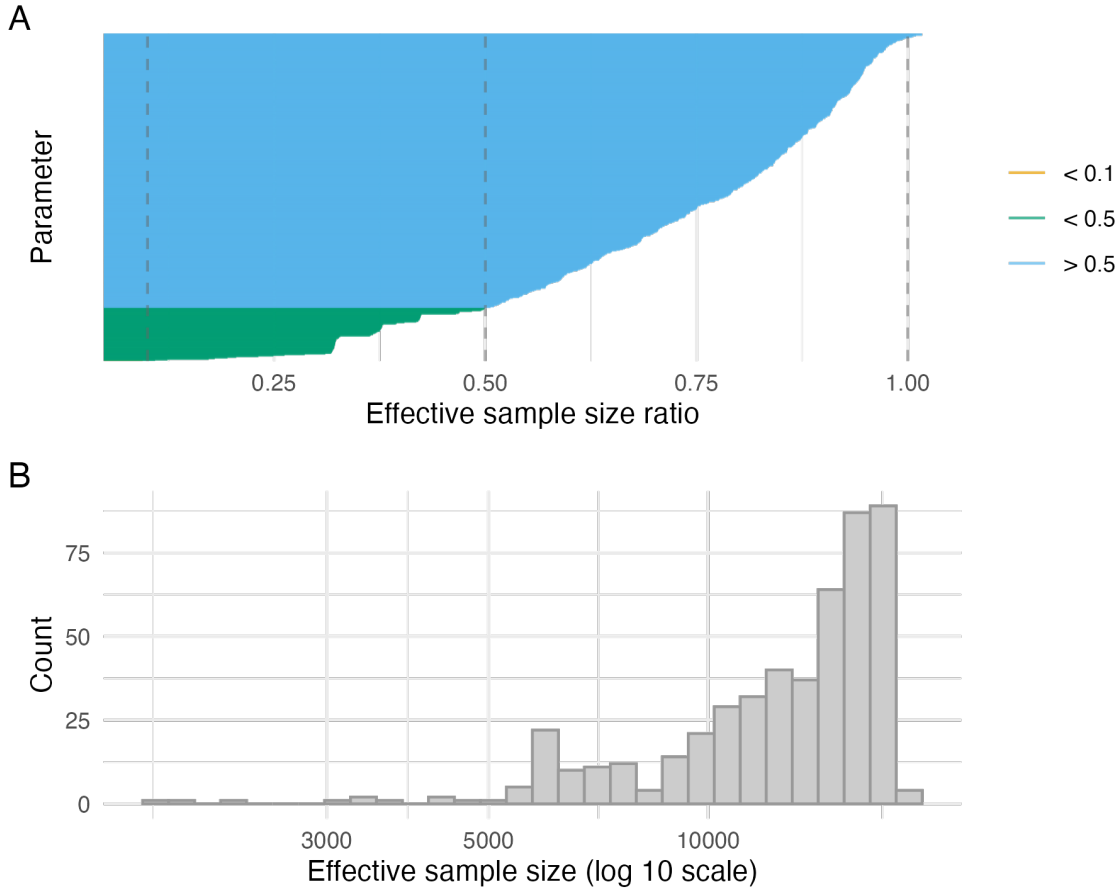

Figure 1: The efficiency of the NUTS, as measured by the ratio of effective sample size to total number of iterations run, was low for some parameters (Panel A). As a result, the number of iterations required for the effective number of samples (mean 13041) to be satisfactory was high (Panel B).

### 2.1 Posterior correlation structure

Correlation structure in the posterior can result in sampler inefficiency. Each of the four pairs of AR1 log standard deviation $\log(\sigma)$ and logit lag-one autocorrelation parameter $\text{logit}(\phi)$ posteriors were positively correlated (mean absolute correlation 0.82, Figure 4). These parameters are partially identifiable as variation can either be explained by high standard deviation and high autocorrelation or low standard deviation and low autocorrelation. On the other hand, the BYM2 log standard deviation $\log(\sigma)$ and logit proportion parameter $\text{logit}(\phi)$ were, as designed, more orthogonal (mean absolute correlation 0.21, Figure 5).

### 2.2 Posterior contraction

The informativeness of data about a parameter can be summarised by the posterior contraction (Schad, Betancourt, and Vasishth 2021) which compares the prior variance $\mathbb{V}_{\text{prior}}(\phi)$ to posterior variance $\mathbb{V}_{\text{post}}(\phi)$ via

$$c(\phi) = 1 - \frac{\mathbb{V}_{\text{prior}}}{\mathbb{V}_{\text{post}}(\phi)}. \tag{16}$$

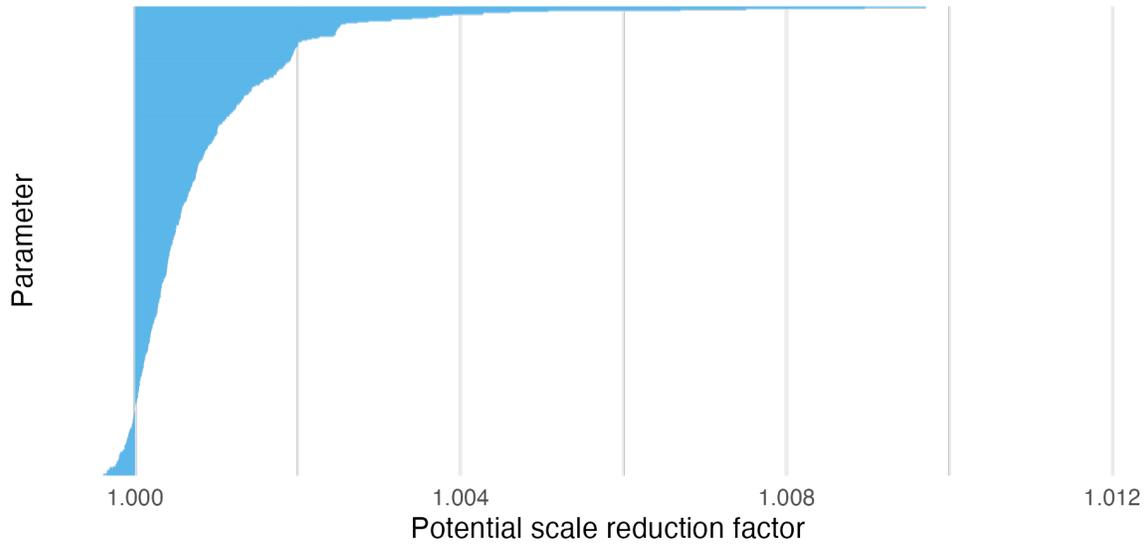

Figure 2: For NUTS run on the Naomi ELGM, the maximum potential scale reduction factor was 1.01, substantially below the value of 1.05 typically used as a cutoff for acceptable chain mixing, indicating that the results are of high quality. Additionally, all $\hat{R}$ values were less than 1.1.

Posterior variances were extracted from NUTS results, and prior variances obtained by simulating from a model with the likelihood component removed (Figure 6). The average posterior contraction was positive for all latent field parameter vectors, and for the majority of hyperparameters (Figure 7). However, for seven hyperparameters the posterior contraction was very close to zero. Furthermore, for some latent field parameter vectors, the average contraction was small. Based on this findings, these parameters may not be identifiable.

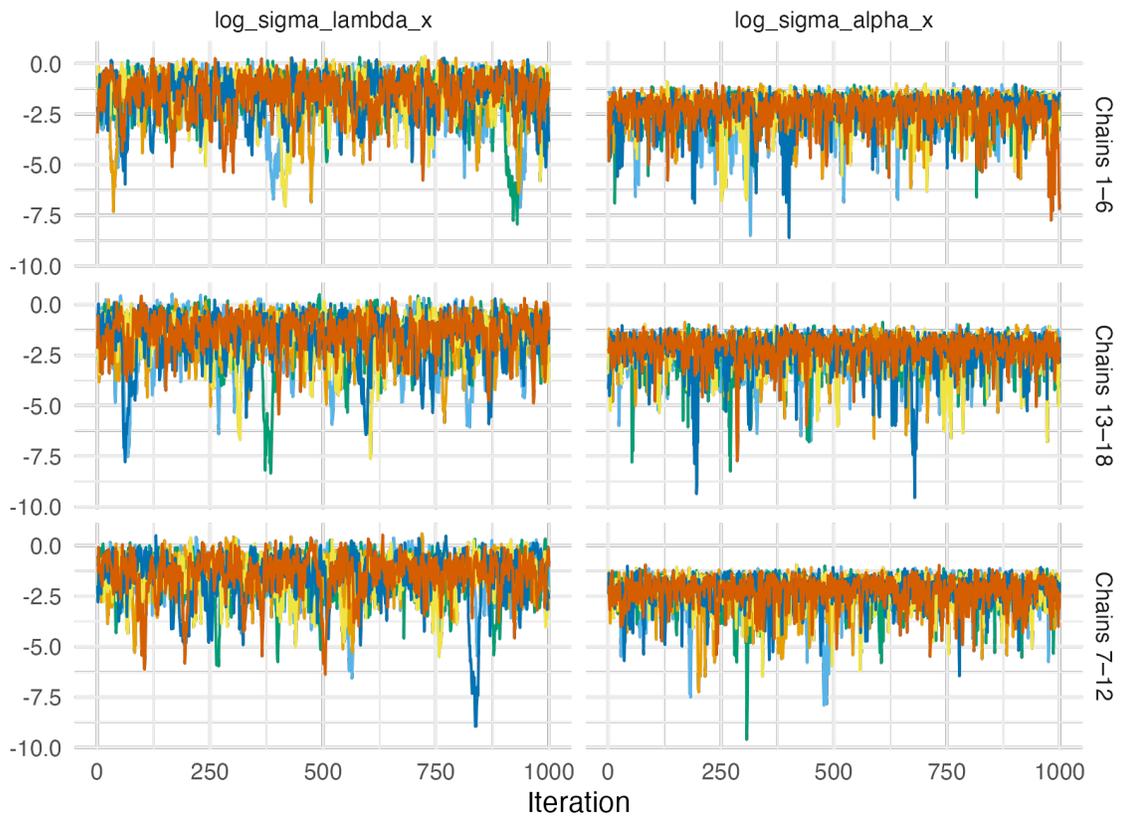

Figure 3: Traceplots for the parameter with the lowest ESS which was `log_sigma_lambda_x` (an ESS of 1696, Panel A) and highest potential scale reduction factor which was `log_sigma_alpha_x` (an $\hat{R}$ of 1.01, Panel B).

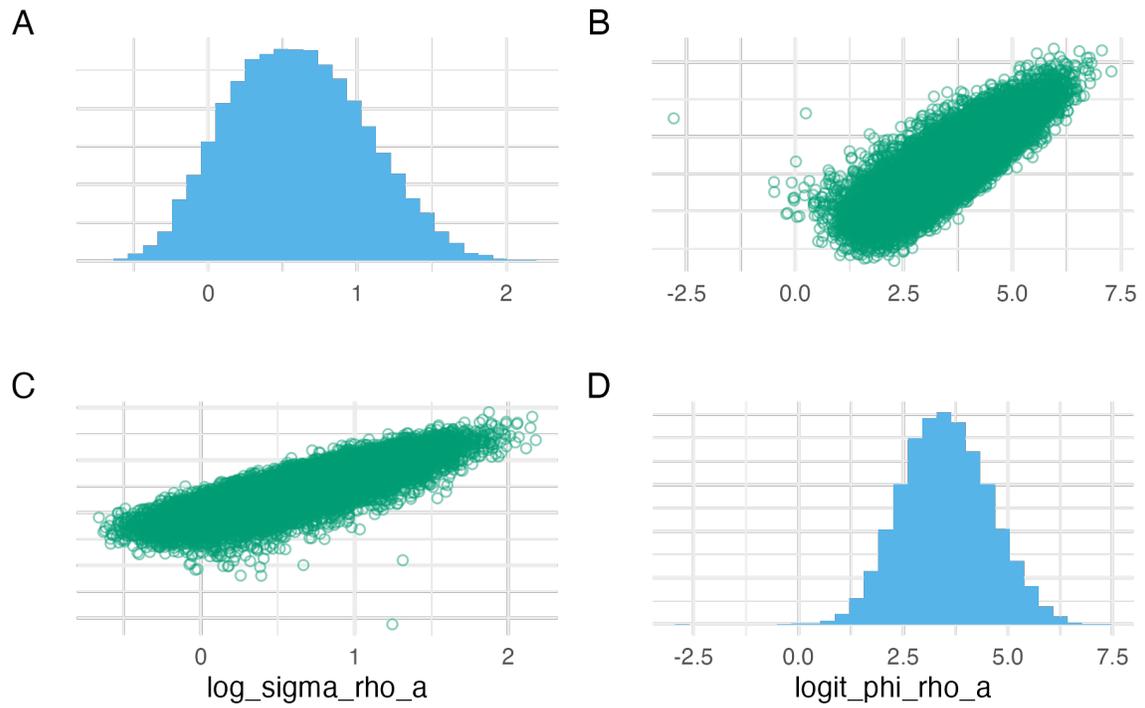

Figure 4: Pairs plots for the parameters $\log(\sigma_A^\rho)$ and $\text{logit}(\phi_A^\rho)$, or `log_sigma_rho_a` and `logit_phi_rho_a` as implemented in code. These parameters are the log standard deviation and logit lag-one correlation parameter of an AR1 process. In the posterior distribution obtained with NUTS, they have a high degree of correlation.

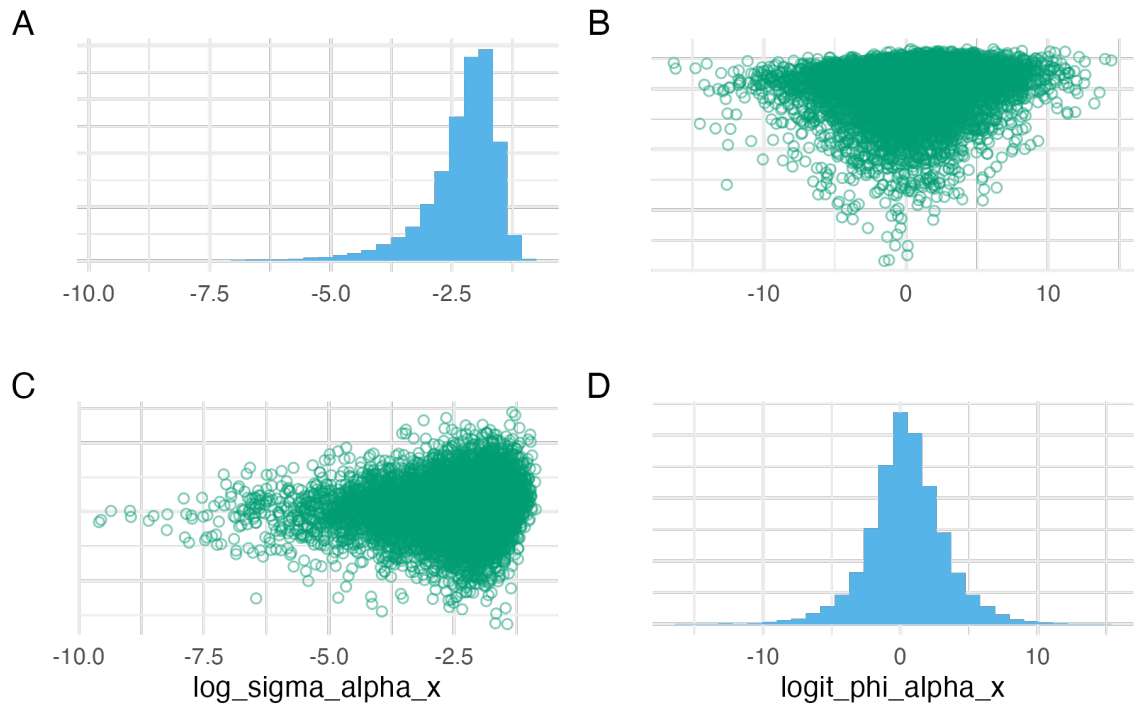

Figure 5: Pairs plots for the parameters $\log(\sigma_X^\alpha)$ and $\text{logit}(\phi_X^\alpha)$, or `log_sigma_alpha_x` and `logit_phi_alpha_x` as implemented in code. These parameters are the log standard deviation and logit BYM2 proportion parameter of a BYM2 process. In the posterior distribution obtained with NUTS, they are close to uncorrelated.

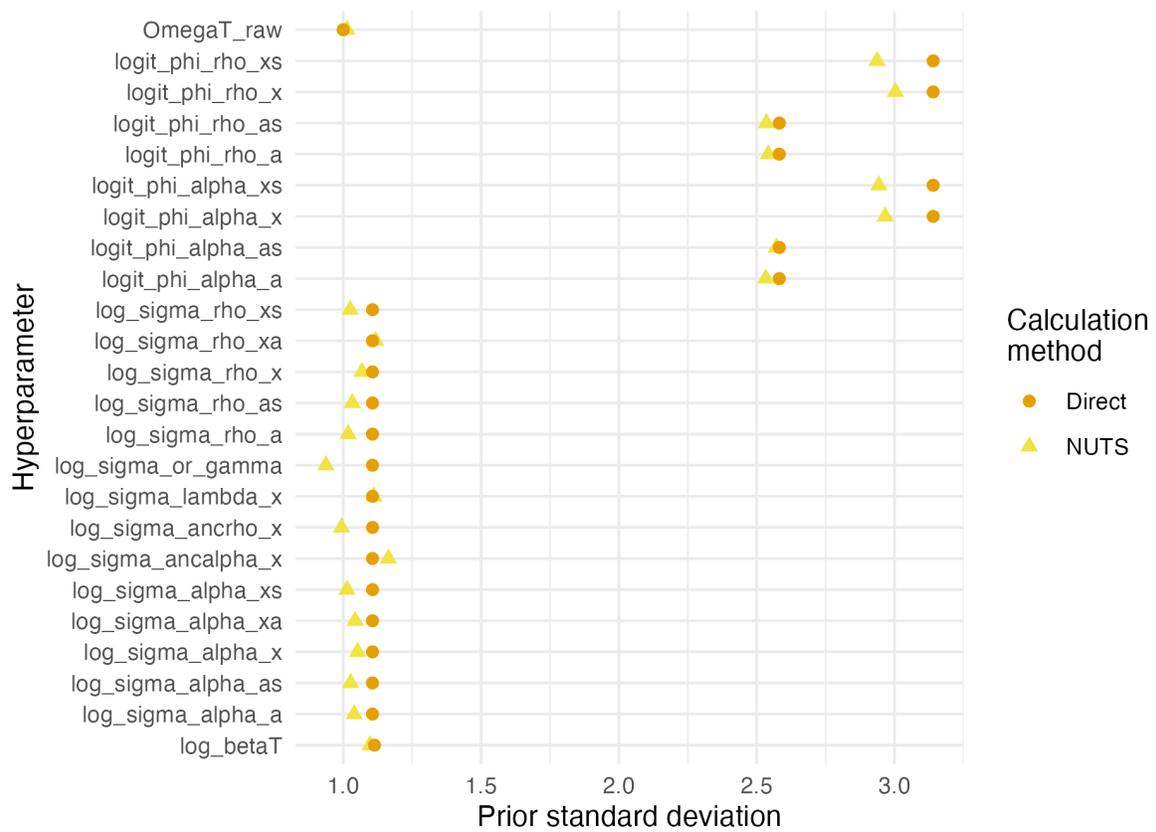

Figure 6: Prior standard deviations were calculated by using NUTS to simulate from the prior distribution. This approach is more convenient than simulating directly from the model, but can lead to inaccuracies.

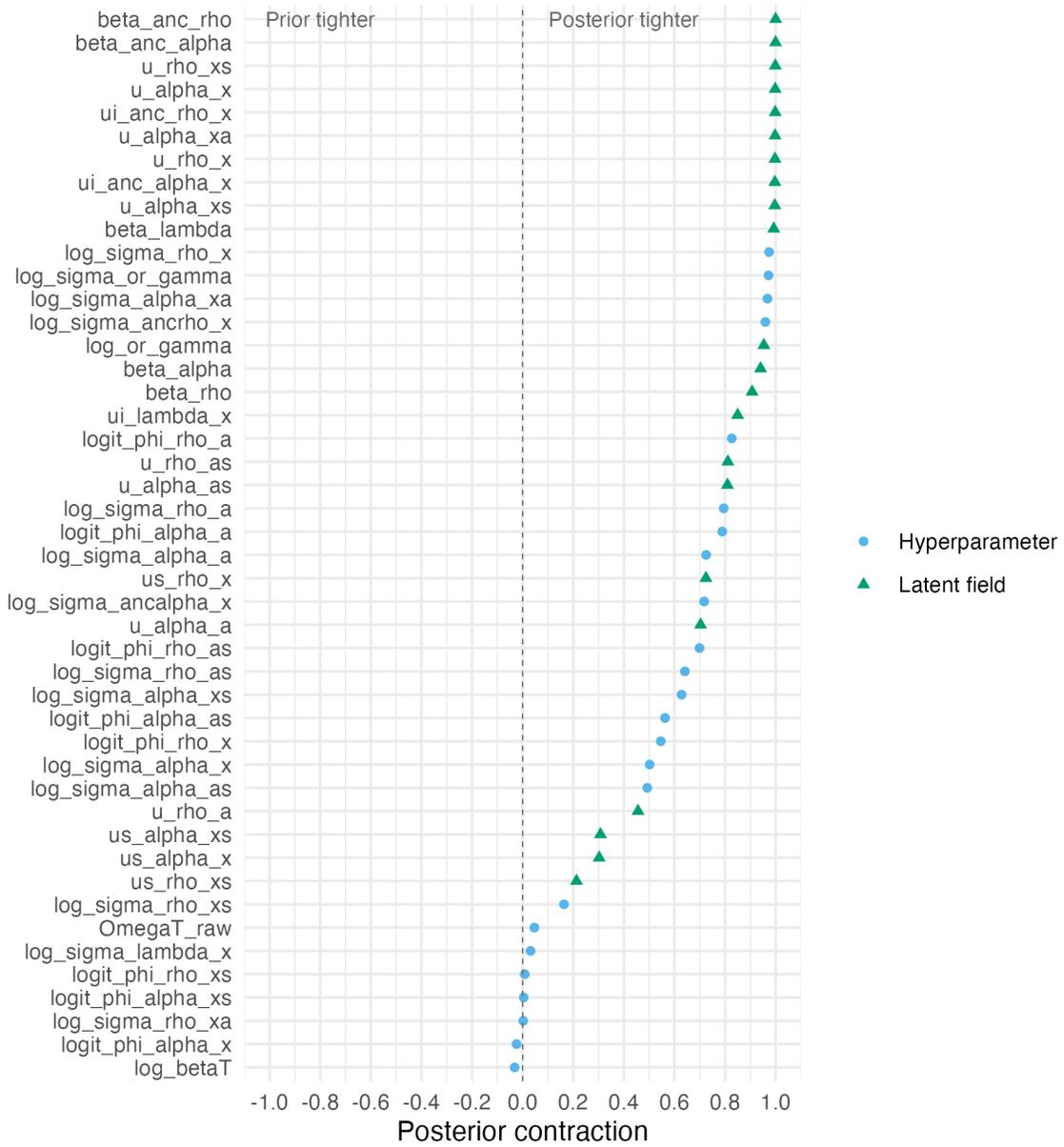

Figure 7: The posterior contraction for each parameter in the model. Values are averaged for parameters of length greater than one. The posterior contraction is zero when the prior distribution and posterior distribution have the same standard deviation. This could indicate that the data is not informative about the parameter. The closer the posterior contraction is to one, the more than the marginal posterior distribution has concentrated about a single point.

# 3 Use of PCA-AGHQ

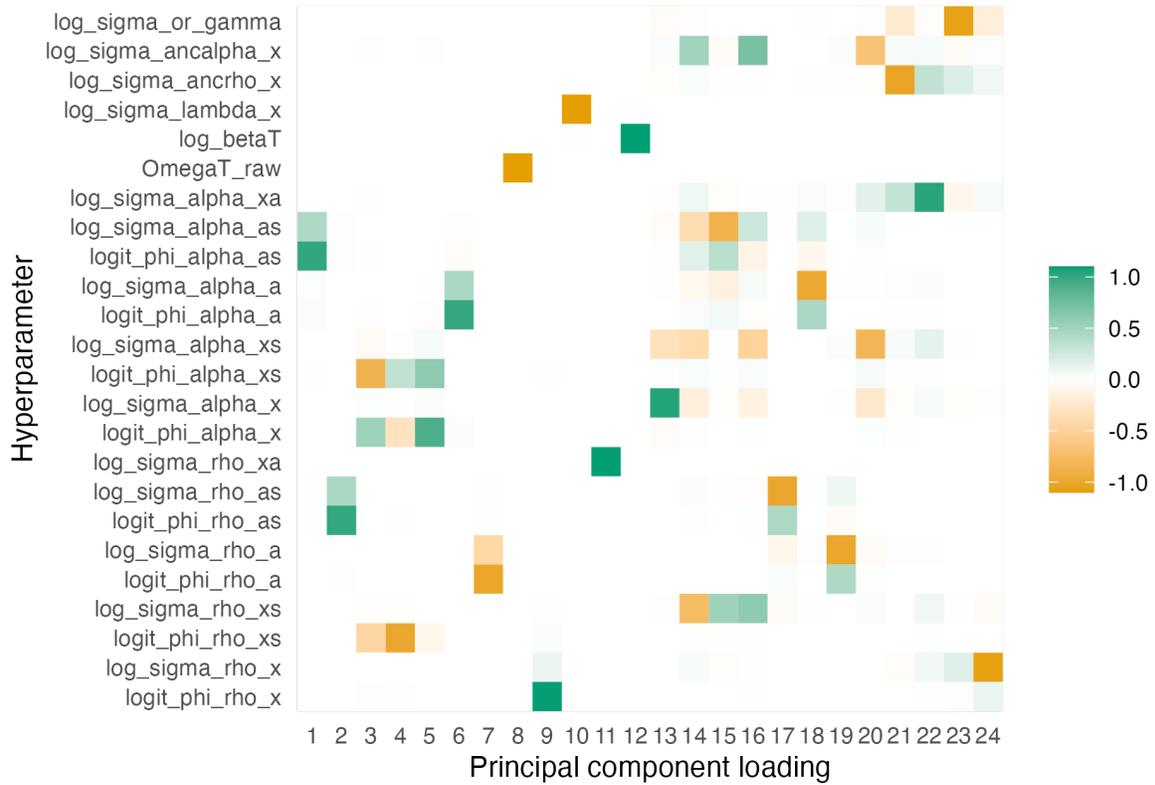

Figure 8: Each principal component loading, obtained by the eigendecomposition of the inverse curvature, gives the direction of maximum variation conditional on inclusion of each previous principal component loading. For example, the first principal component loading is a sum of `log_sigma_alpha_as` and `logit_phi_alpha_as`.

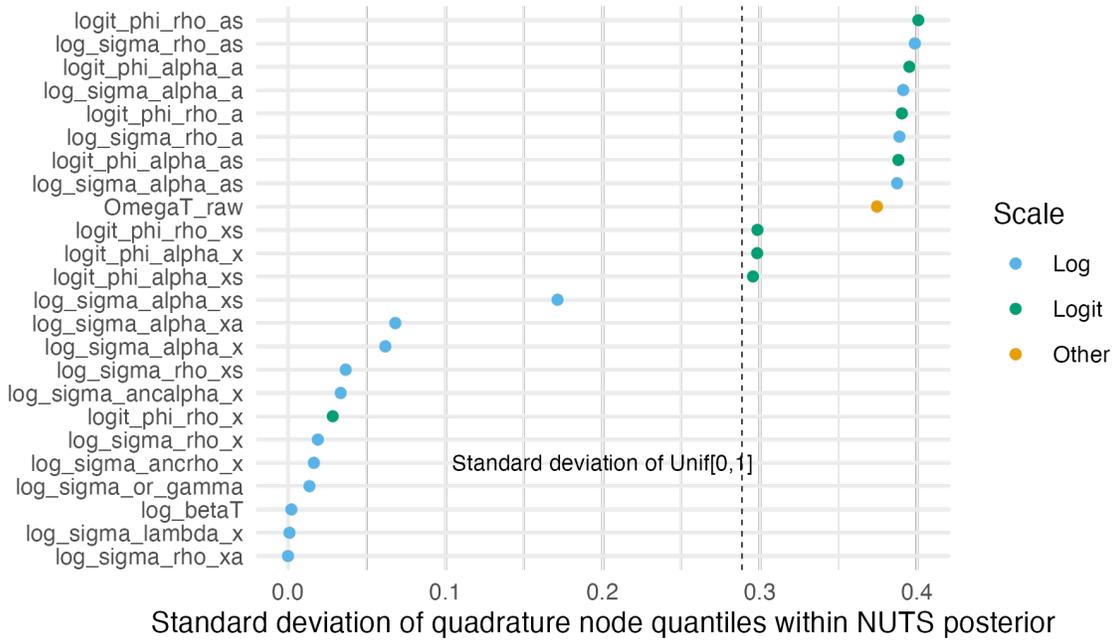

Figure 9: The standard deviation of the quadrature nodes can be used as a measure of coverage of the posterior marginal distribution. Nodes spaced evenly within the marginal distribution would be expected to uniformly distributed quantile, corresponding to a standard deviation of 0.2894, shown as a dashed line.

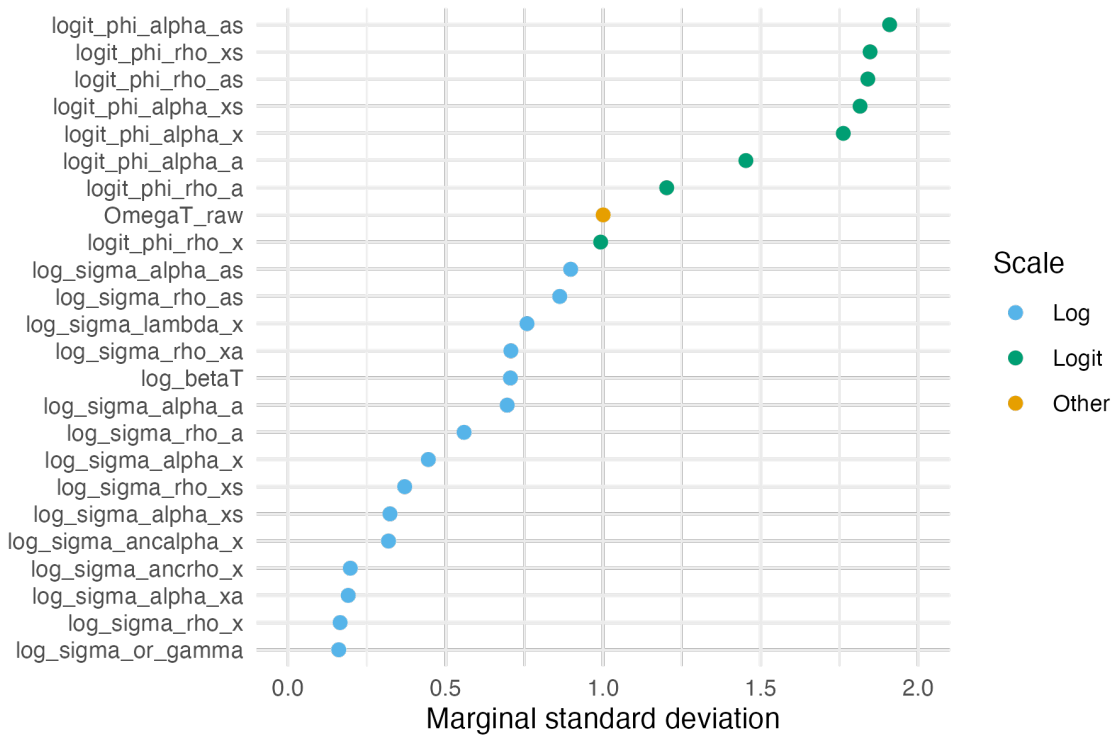

Figure 10: The estimated posterior marginal standard deviation of each hyperparameter varied substantially based on its scale, either logarithmic or logistic.

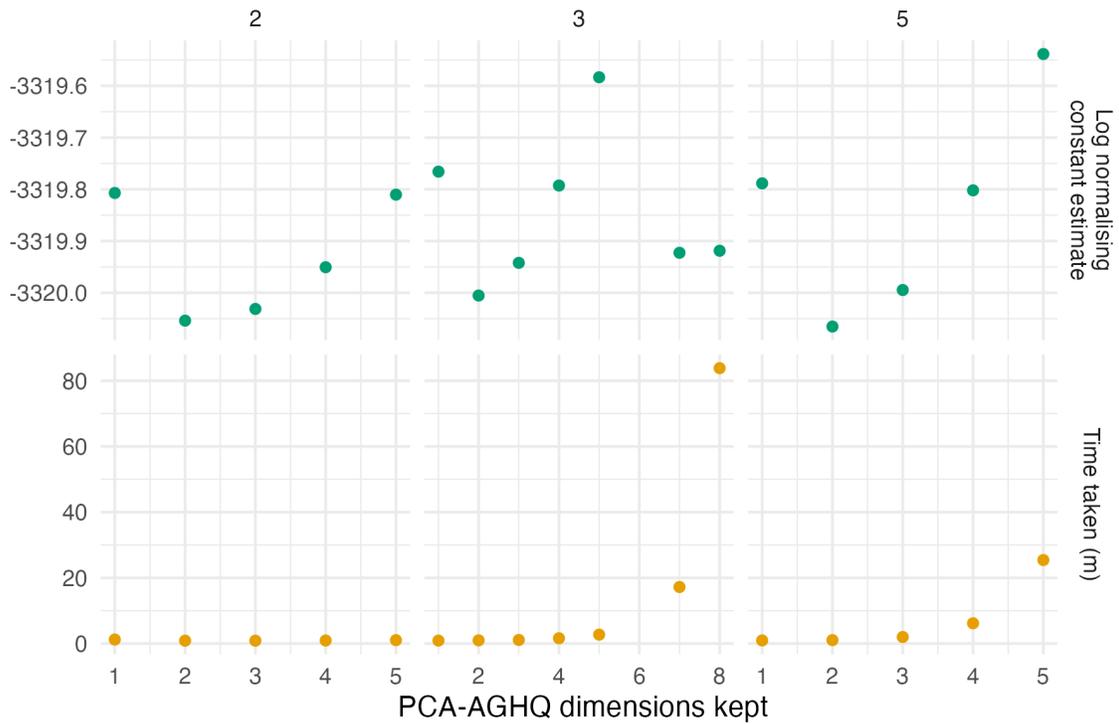

Figure 11: The logarithm of the normalising constant estimated using PCA-AGHQ and a range of possible values of $k = 2, 3, 5$ and $s \leq 8$. Using this range of settings, there was not convergence of the logarithm of the normalising constant estimate. The time taken by GPCA-AGHQ increases exponentially with number of PCA-AGHQ dimensions kept.

# 4 Inference comparison

## 4.1 Point estimates

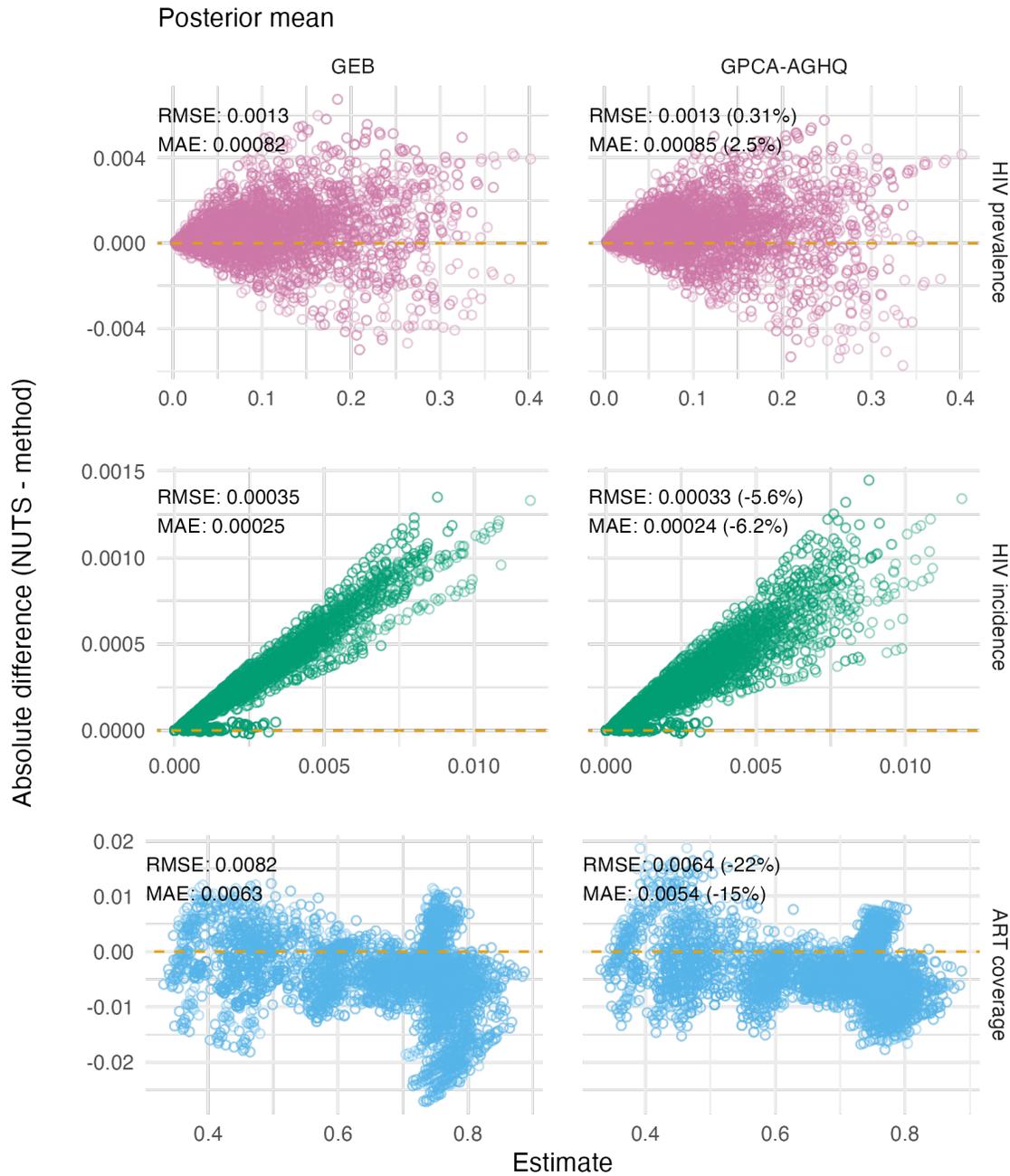

Figure 12: Differences in Naomi model output posterior means as estimated by GEB and GPCA-AGHQ compared to NUTS. Each point is an estimate of the indicator for a particular strata. In all cases, error is reduced by GPCA-AGHQ, most of all for ART coverage.

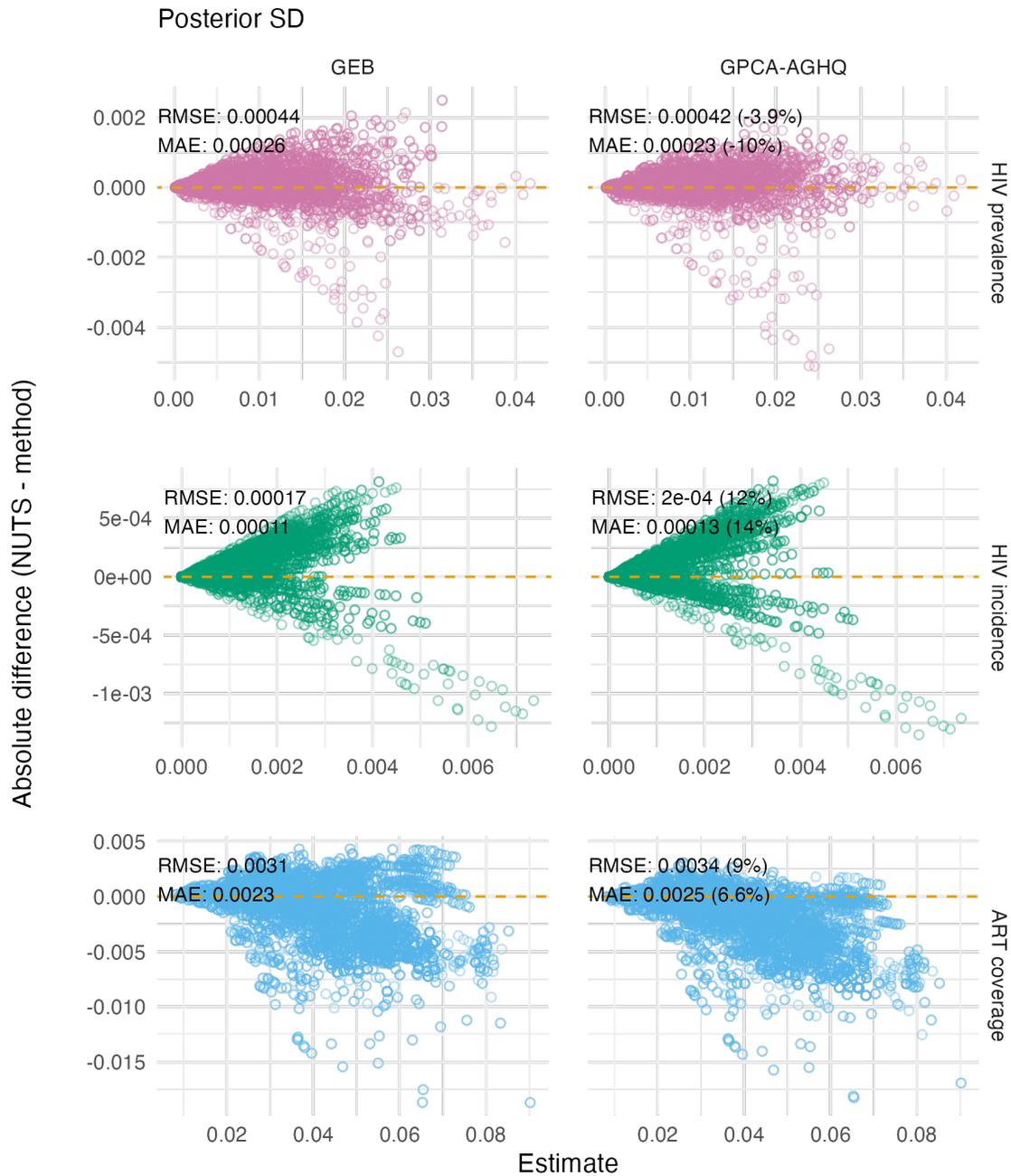

Figure 13: Differences in Naomi model output posterior standard deviations as estimated by GEB and GPCA-AGHQ compared to NUTS. Each point is an estimate of the indicator for a particular strata. Error is increased by GPCA-AGHQ for ART coverage and HIV incidence, and reduced for HIV prevalence.

## 4.2 Distributional quantities

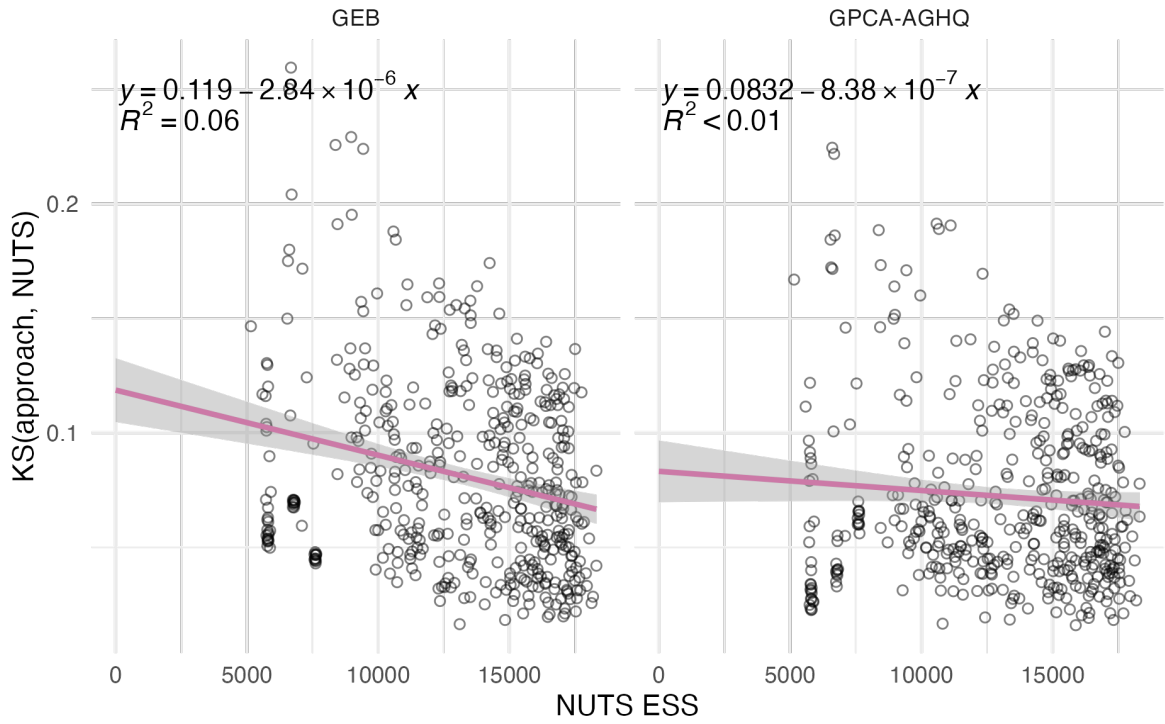

Figure 14: The Kolmogorov-Smirnov (KS) test statistic for each latent field parameter is correlated with the effective sample size (ESS) from NUTS, for both GEB and GPCA-AGHQ. This may be because parameters which are harder to estimate with INLA-like methods also have posterior distributions which are more difficult to sample from. Alternatively, it may be that high KS values are caused by inaccurate NUTS estimates generated by limited effective samples.